\documentclass[pdflatex,sn-mathphys-num]{sn-jnl}

\newlength{\figurewidth}
\usepackage{graphicx}%
\usepackage{multirow}%
\usepackage{amsmath,amssymb,amsfonts}%
\usepackage{amsthm}%
\usepackage{mathrsfs}%
\usepackage[title]{appendix}%
\usepackage{xcolor}%
\usepackage{textcomp}%
\usepackage{manyfoot}%
\usepackage{booktabs}%
\usepackage{longtable}
\usepackage{algorithm}%
\usepackage{algorithmicx}%
\usepackage{algpseudocode}%
\usepackage{listings}%
\usepackage{siunitx}%
\usepackage{textcomp}

\usepackage[utf8]{inputenc}

\begin{document}

\addtolength{\textheight}{0.8in}
\setlength{\footskip}{0.1in}

\title[Pulse Shaping Increases Efficiency in Pulsed Plasma Accelerators]
{Pulse Shaping Increases Efficiency in Pulsed Plasma Accelerators}

\author[1]{\fnm{Patrick W.} \sur{Schools}}

\author[1]{\fnm{Mammadbaghir} \sur{Baghirzade}}

\author[1]{\fnm{Ryan} \sur{Heiser}}

\author[1]{\fnm{Adrian} \sur{Woodley}}

\author[1]{\fnm{Laxminarayan L.} \sur{Raja}}

\author*[1,2]{\fnm{Thomas C.} \sur{Underwood}}
\email{thomas.underwood@utexas.edu}

\affil*[1]{\orgdiv{Department of Aerospace Engineering and Engineering Mechanics},
\orgname{The University of Texas at Austin},
\orgaddress{\city{Austin}, \state{Texas}, \postcode{78712}, \country{USA}}}

\affil[2]{\orgdiv{Texas Materials Institute},
\orgname{The University of Texas at Austin},
\orgaddress{\city{Austin}, \state{Texas}, \postcode{78712}, \country{USA}}}

\abstract{
The performance of a gas-fed pulsed electromagnetic thruster is governed by the ability to deposit electrical energy while propellant is available for acceleration.  Current pulsed-power systems require tradeoffs between high-current discharges that produce high exhaust velocities and longer pulses that overlap the energy deposition with more of the gas injection. This limited control restricts the specific impulse and mass utilization of these thrusters. This work introduces programmable pulse shaping as a method to increase control over energy deposition and expand the accessible operating space. We use solid-state integrated power modules to vary the discharge delay, pulse width, peak current, and the shape of the current waveform as propellant is injected. Experiments varying pulse widths from 30 to 500 $\mu$s and peak currents from 3 kA to 16 kA show that short, high-current pulses produce higher exhaust velocities and greater impulse bits than longer, lower-current pulses at comparable discharge energy. The switches also enable multiple discharges of arbitrary positioning and duration during a single gas injection. This micro-burst operation is shown to increase specific impulse in air by 278\% from 840 to 3177 s through improved propellant utilization. This same pulse shaping ability is found to increase thrust efficiency from 0.2\% in single-shot operation to 3.3\% in micro-burst operation while operating with air. 
}


\maketitle
\section{Introduction}

Emerging missions require spacecraft to travel farther, operate longer, and carry larger scientific or commercial payloads without increasing launch mass. \cite{Lev2019ThePropulsion,Benkhoff2021,Johnson2002} These missions require propulsion systems that can achieve high specific impulse ($I_{\mathrm{sp}}$) to reduce the propellant that is needed to achieve large cumulative changes in velocity, preserve mass for payload, and extend operational lifetime. \cite{Benkhoff2021,Johnson2002,Woodley2024RequirementsOrbits} Similar requirements arise for spacecraft operating in very low Earth orbit (VLEO), where continuous drag compensation requires both efficient propulsion and a continuous propellant supply. \cite{Crandall2022Air-breathingAnalysis} Air-breathing electric propulsion operates in this regime by harvesting atmospheric gas as a self-replenishing propellant source. However, the amount of propellant that can be collected is limited by gas-dynamic constraints. At an altitude of 220 km, for example, a representative spacecraft with a collection efficiency of 50\% can harvest only approximately 0.1-1 mg/s, depending on the ratio of intake area to frontal area. \cite{Woodley2024RequirementsOrbits} The limited propellant flow and mixed composition of molecular and atomic species require propulsion systems that are capable of producing high exhaust velocities while converting the available propellant into thrust efficiently. Among existing electric propulsion architectures, electrostatic thrusters can achieve high specific impulse but are limited fundamentally in the ion current that can be extracted by space-charge effects. \cite{Lev2019ThePropulsion,Chabert2012GlobalCoil,Taccogna2023PlasmaAlgorithms} This constraint makes it difficult to scale electrostatic acceleration toward higher $I_{\mathrm{sp}}$, thrust efficiency ($\eta$), and thrust density simultaneously, particularly when molecular atmospheric propellants introduce additional ionization and dissociation losses. \cite{Crandall2024,Marchioni2021ExtendedPropulsion} These limitations motivate the development of alternative acceleration mechanisms that are capable of achieving high thrust efficiency with minimal amounts of propellant.

Electromagnetic thrusters avoid these scaling constraints by accelerating a plasma through the Lorentz force rather than separating and accelerating an ion beam through an electrostatic potential. \cite{BURTON1981,Choueiri2001,Krulle1998TechnologyPropulsion} This mechanism can be scaled through the discharge current and magnetic field to produce higher exhaust velocities and, in theory, higher $I_{\mathrm{sp}}$. \cite{Choueiri1998ScalingThrusters,Choueiri2001,LaPointe2004} How this acceleration is scaled depends in part on the source of the magnetic field. Self-field devices generate this field through the discharge current itself, while applied-field devices use an externally generated field to supplement or replace the self-field. \cite{Krulle1998TechnologyPropulsion} Regardless of the magnetic field source, the current scaling that is necessary to achieve high thrust and $I_{\mathrm{sp}}$ requires high-power operation that can both ablate electrodes and melt them. \cite{Gallimore1993AnodeThrusters,Choueiri1998ScalingThrusters,Zuin2004CriticalThruster} Because these electrical and thermal loads must be sustained continuously, the available spacecraft power and allowable electrode heat load constrain the discharge current that can be maintained. This lower discharge current limits the attainable electromagnetic force and, consequently, the exhaust velocity and $I_{\mathrm{sp}}$.

\begin{figure}[bp!]
    \centering
    \includegraphics[width=\figurewidth]{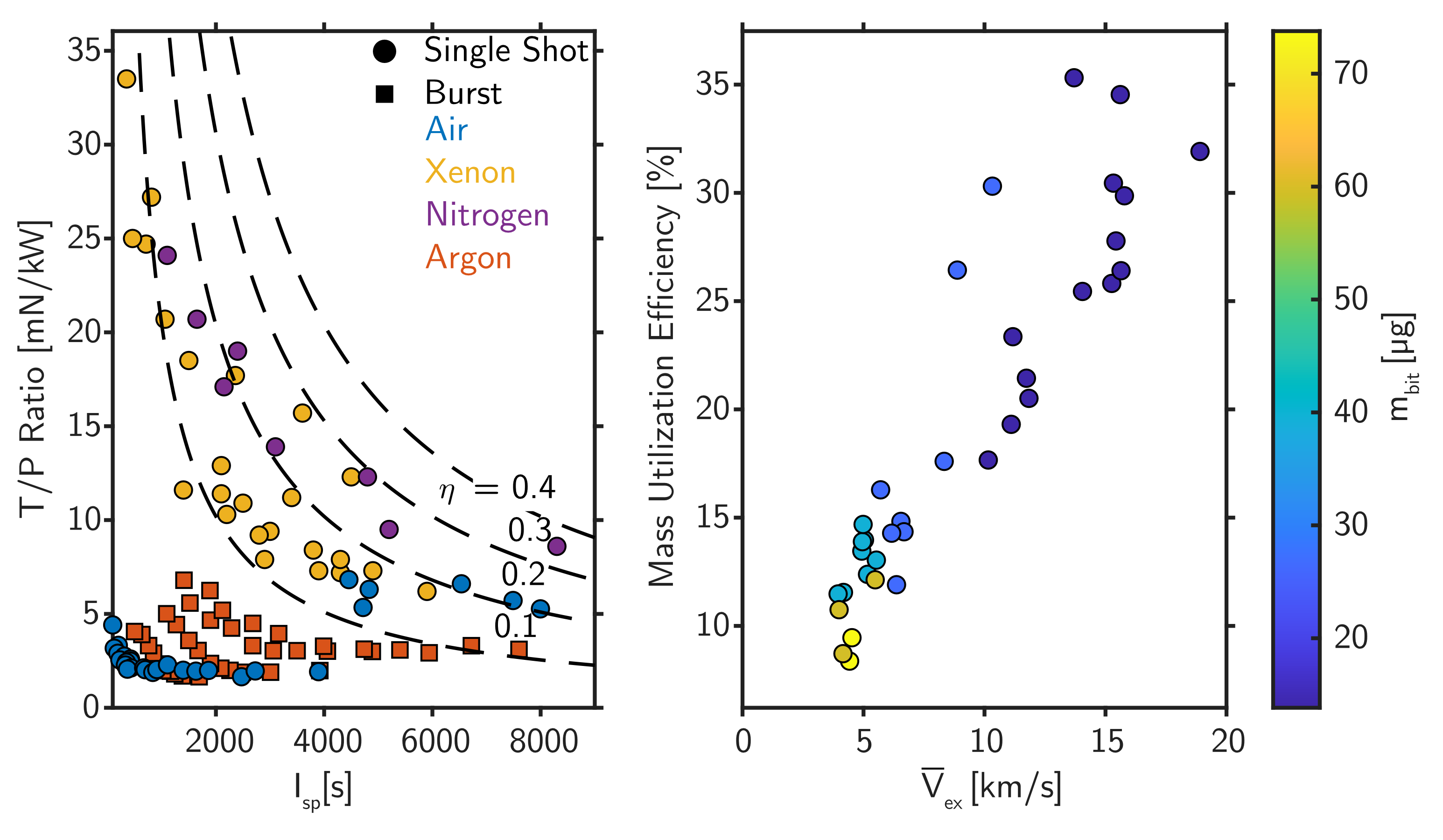}
    \caption{Performance metrics for self-field gas-fed pulsed plasma thrusters compiled from previous studies. \cite{Ziemer2001,Subhankar2024,Woodley2025b} a) Thrust-to-power ratio ($T/P$), specific impulse ($I_{\mathrm{sp}}$), and the corresponding thrust efficiency, $\eta=(g_0/2)I_{\mathrm{sp}}(T/P)$, for GFPPTs and related pulsed electromagnetic accelerators using different propellants. b) Mass utilization efficiency versus average exhaust velocity, $\overline{V}_{\mathrm{ex}}$, for various injected mass bits.}
    \label{fig:current_literature}
\end{figure}

Pulsed electromagnetic thrusters address the high continuous-power requirement of steady-state electromagnetic acceleration by storing energy between firings and releasing it in brief, high-current discharges. \cite{Ziemer2001,Subhankar2024,Choueiri2000,ZiemerAHalf-Century,Toki2000On-orbitSystem} These high-current discharges generate high exhaust velocities and have been shown to achieve high $I_{\mathrm{sp}}$ in some operating configurations where gas is injected (Fig.~\ref{fig:current_literature}a). Examples include ablative pulsed plasma thrusters, which vaporize and ionize material from a solid propellant surface, and gas-fed devices, including gas-fed pulsed plasma thrusters (GFPPTs) and pulsed magnetoplasmadynamic thrusters (pulsed MPDs). \cite{Molina-CabreraPPulsedClassification} Ablative thrusters benefit from having propellant continuously available at the discharge surface, but their performance can be affected by incomplete ionization, particulate ejection, and changes in the propellant surface over time. \cite{Keidar2004,Zhang2019AThrusters,Huang2020,Huang2015StudyThruster} Gas-fed devices instead introduce gaseous propellant independently of the discharge as a means to minimize the ablation of thruster components. \cite{Woodley2026ConnectingThrusters}

Gas-fed pulsed electromagnetic thrusters introduce additional challenges because the deposition of electrical energy must be coordinated with propellant addition. While these devices can produce peak plume velocities exceeding 60 km/s, many studies report their average exhaust velocities, $\overline{V}_{\mathrm{ex}}$, remain below 20 km/s (Fig.~\ref{fig:current_literature}b). \cite{Woodley2025b} This discrepancy indicates that only a fraction of the injected propellant is entrained within the discharge, while the remaining mass does not reach the accelerator volume before the discharge ends. Experiments have confirmed that mass utilization in GFPPTs that operate with short pulse widths of 10-50 $\mu$s can be as low as 9\% during single discharge pulses (Fig.~\ref{fig:current_literature}b). \cite{Woodley2025b} Poor mass utilization in these conditions reflects a mismatch between the timescales of gas injection and electrical energy deposition. \cite{Loebner2015,Choueiri2000,ZiemerAHalf-Century} Among these conditions, short-duration discharges have been shown to be the most sensitive to the gas dynamics of propellant filling the accelerator volume. \cite{Woodley2025b,Underwood2021,Loebner2015Branch} If discharge energy is held constant, increasing the pulse duration of these devices improves temporal overlap with the injected propellant but reduces the peak current and weakens the peak electromagnetic forces that can be achieved. The central challenge is to develop mechanisms to maintain high peak current operation over a longer time period to accelerate a larger fraction of the injected propellant. \cite{Ziemer2001}

The lack of control over energy addition confines gas-fed pulsed thrusters to a limited range of operating conditions. This limitation has motivated the development of pulse-forming networks (PFNs) and switched pulsed-power systems to better match the gas and energy timescales and control discharge initiation. \cite{Toki2000On-orbitSystem,Polzin2020State-of-the-artThrusters} In PFNs, the capacitance and inductance determine the peak current, pulse duration, and current waveform, such that changing these discharge characteristics requires reconfiguring the pulsed-power hardware. \cite{Castillo1991,Cheng1971ApplicationThruster,York1993DiagnosticsNozzle,Zimmerman2025PulsedOrbit} Switched systems have also been used to vary the discharge initiation time, which determines the initial propellant distribution and can alter the acceleration structure and exhaust velocity of the resulting ionization wave. \cite{Loebner2015Branch} However, this approach controls only the conditions at breakdown. Existing implementations using gaseous and solid-state switches have been generally limited to a single high-current switching event per discharge, and previous solid-state supplies have operated similarly as closing switches that release all stored energy in a single pulse and must recharge before producing another discharge. \cite{Winands2005LongApplications,Ziemer1999IsPerformance,Laya2026AGun,Ziemer1997PerformanceThruster} These approaches provide control over the initial propellant condition at breakdown but little control over how energy is delivered subsequently as the propellant evolves. A remaining challenge is to develop schemes to control the timing, duration, and shape of waveforms to match the gas dynamics of thrusters without requiring new hardware for each operating configuration. \cite{Promislow2022OperationRates,Polzin2020State-of-the-artThrusters}

This work uses programmable solid-state switches to control energy addition directly through high-frequency pulse width modulation (PWM). \cite{Ziemba2011EHTIGBT,Miller2013EHTIPM} Previously, pulse widths, peak currents, pulse delays, and current waveform shapes have been constrained by the circuitry of a power supply. Programmable pulse shaping enables the variation of these pulse characteristics and breaks constraints between timescales that govern propellant gas dynamics and plasma evolution during the acceleration process. This control enables energy to be deposited selectively as the plasma is evolving or distributing energy optimally over longer periods where propellant is available. We demonstrate this through studies on the effect of pulse width, pulse energy, peak current, and pulse shape to compare the performance of electromagnetic acceleration in different operating regimes. This improved control also enables the application of multiple short, high-current discharges during a single gas injection event, removing the need to choose between high peak current and long pulse widths that overlap with more of the propellant availability for a given average input power. We also demonstrate how programmable discharge timing sets the initial propellant loading in the accelerator. \cite{Loebner2015,Underwood2021} These timing changes result in differences in acceleration structures and performance. \cite{Woodley2025b} By controlling discharge delay, peak current, pulse duration, pulse count, interpulse spacing, and pulse-to-pulse energy allocation, pulse shaping allows a single accelerator to access and evaluate operating characteristics that would otherwise require a collection of different pulsed-power supplies. By uncoupling the peak operating current from the energy addition timescale, this expanded design space provides a pathway to improve the $I_{\mathrm{sp}}$ of gas-fed pulsed thrusters.

\section{The Pulsed Electromagnetic Thruster Design Space}

\begin{figure}[tbp!]
    \centering
    \includegraphics[width=\figurewidth]{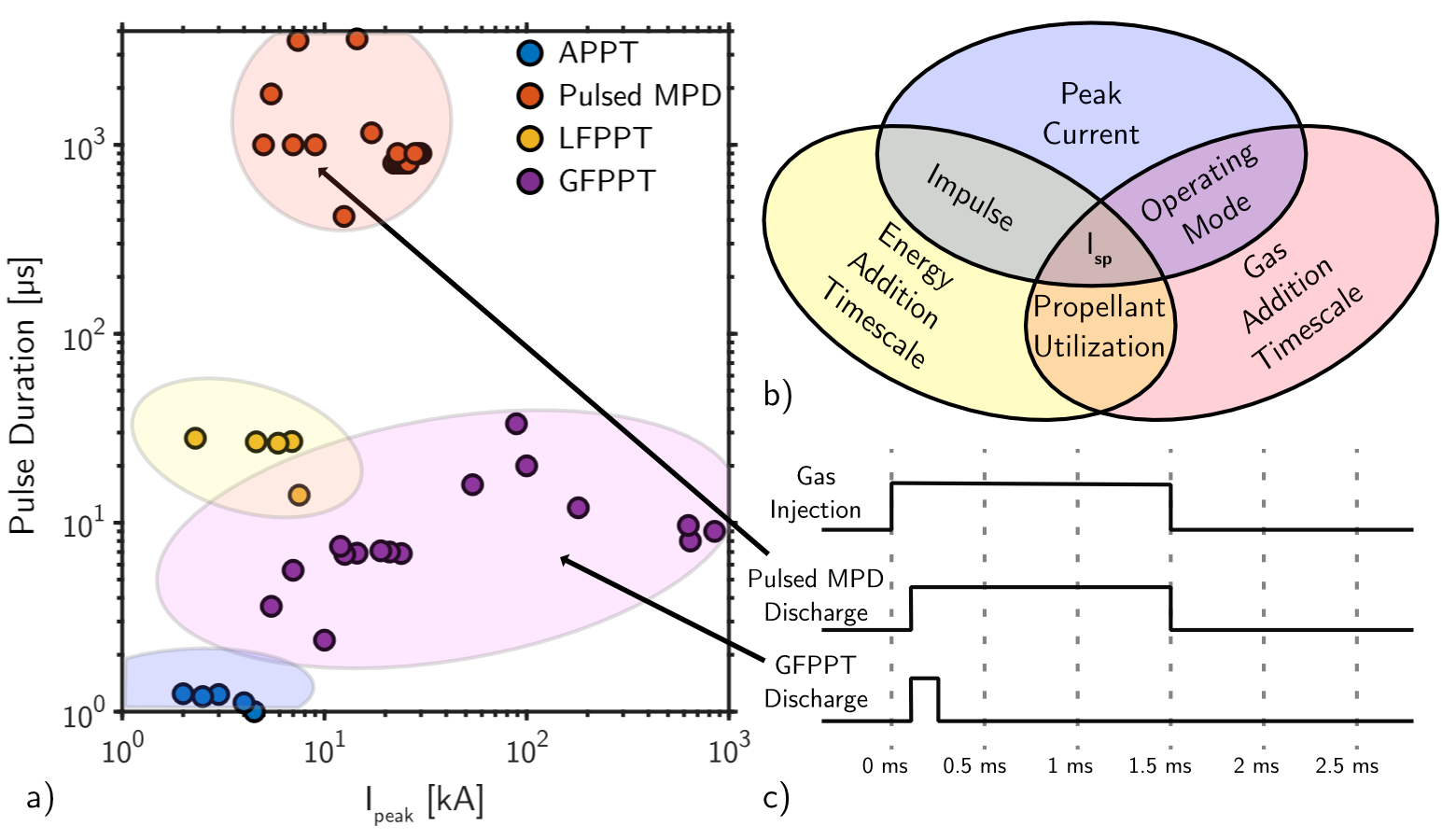}
    \caption{The distinct operating regimes of various pulsed electromagnetic thrusters. a) Pulsed electromagnetic devices occupy different regions defined primarily by pulse duration and peak current. \cite{Domonkos1995,LaPointe2004,Choueiri2001,Zimmerman2025PulsedOrbit,Ducati1971InvestigationJets,Organski2025,Ziemer2000,Underwood2021,Poehlmann2010CurrentMode,Underwood2017,Cheng1971ApplicationThruster,Hauze1992EffectPerformance,Woodall1985ObservationDeflagration,Huang2015StudyThruster,Zhang2019AThrusters,Huang2020} b) The design space for gas-fed pulsed electromagnetic thrusters includes the gas addition timescale, energy addition timescale, and peak current. These variables influence impulse, propellant utilization, thrust efficiency, and acceleration mode. c) Timing diagrams illustrating the characteristic energy addition timescales of GFPPTs and pulsed MPDs relative to gas injection. GFPPTs deposit energy over a small fraction of the gas-injection event, whereas pulsed MPDs operate over timescales closer to the gas addition timescale.}
    \label{fig:devicecomparison}
\end{figure}

The design space of gas-fed pulsed electromagnetic thrusters is governed primarily by the gas addition timescale, $\tau_G$, and the energy addition timescale, $\tau_E$ (Fig.~\ref{fig:devicecomparison}b). The gas addition timescale determines the propellant loading encountered by the discharge, while the energy addition timescale determines how long the Lorentz force is sustained. Additional effects emerge when $\tau_E$ approaches the accelerator flow-through timescale, $\tau_C = L/V$, where $L$ is the accelerator length and $V$ is the plasma velocity. When $\tau_E\sim\tau_C$, a larger fraction of the discharge occurs before the initially loaded propellant exits the accelerator. This makes thruster performance more sensitive to the initial propellant distribution and the resulting magnetohydrodynamic operating modes that the thruster produces initially. \cite{Woodley2025b,Underwood2021,Loebner2015Branch} The influence of $\tau_E$ is further complicated in conventional pulsed-power systems by its coupling to the peak current. Short energy addition timescales produce high-current discharges, while depositing the same amount of energy over a longer timescale reduces the peak current and thus the achievable peak exhaust velocity. Programmable pulse shaping expands this design space by decoupling peak current from the duration and timing of energy addition. This allows the influence of each parameter to be investigated independently.

The relationship between plasma and propellant gas dynamics introduces fundamental tradeoffs in thruster design between exhaust velocity and propellant utilization. Because of these tradeoffs, two opposing operating regimes have emerged. Pulsed MPDs use PFNs to extend energy addition over hundreds of microseconds to milliseconds (Fig.~\ref{fig:devicecomparison}a,c). \cite{Castillo1991,York1993DiagnosticsNozzle,Zimmerman2025PulsedOrbit} These longer discharges match $\tau_G$ but weaken the Lorentz force for a fixed discharge energy. GFPPTs and pulsed plasma accelerators instead release stored energy through a low-inductance circuit over tens of microseconds and can exceed 100 kA in peak current. \cite{Cheng1971ApplicationThruster,Hauze1992EffectPerformance} Many of these devices operate in a Paschen breakdown configuration such that the discharge begins only when the pressure in the accelerator can support breakdown. \cite{Ziemer2001,ZiemerAHalf-Century} Any mass that enters after this short discharge remains unutilized (Fig.~\ref{fig:devicecomparison}a,c). Some devices implement switches to delay the discharge initiation. However, too large of a delay will result in increased upstream propellant mass and can change the magnetohydrodynamic operating mode of the thruster. \cite{Woodley2025b,Underwood2021}

The initial propellant distribution establishes the conditions for the discharge and determines how the current-driven ionization wave develops as it propagates through the accelerator. \cite{Loebner2015Branch,Underwood2021} Two distinct magnetohydrodynamic flow structures can emerge depending on the initial propellant loading. When breakdown occurs in a densely filled accelerator, the accelerating plasma encounters upstream propellant and compresses it into a dense current sheet. Joule heating within this current sheet increases the electrical conductivity, which concentrates the current locally, and further drives heating and narrowing of the sheet. This positive feedback produces a magneto-detonation wave, in which a current sheet propagates through the upstream propellant in a snowplow-like manner. \cite{Woodley2025b} The magneto-detonation exits the accelerator at relatively low exhaust velocity and is followed by a magneto-deflagration wave, an expanding current-driven plasma jet that accelerates the remaining propellant to substantially higher velocities. In contrast, when breakdown occurs in a sparsely filled accelerator, the plasma encounters little upstream propellant, and the discharge evolves by expanding into vacuum at high velocity and forms a magneto-deflagration wave directly. \cite{Loebner2015Branch}

\subsection{Benefits of the GFPPT Regime}

The primary advantage of the GFPPT regime is its ability to achieve high exhaust velocity and specific impulse by increasing the ratio of peak power to average power. The effective exhaust velocity produced by a pulsed plasma accelerator is given by $V_{\mathrm{ex}}=I_{\mathrm{bit}}/m_{\mathrm{bit}}$, where $I_{\mathrm{bit}}$ is the impulse bit and $m_{\mathrm{bit}}$ is the injected propellant mass. Increasing $V_{\mathrm{ex}}$ requires producing more impulse from each unit of injected mass. In practice, the attainable combination of discharge current, pulse duration, and propellant mass is constrained by the available average input power. Sustaining a high-current discharge for the time required to process a large mass bit demands a large pulse energy. At a fixed repetition rate, this requires a large average power, which is constrained by the thermal loading and degradation within a thruster. The GFPPT regime instead concentrates the available pulse energy into a smaller fraction of the propellant mass by applying higher currents over a shorter pulse duration. By then spacing out these high-current pulses, this produces a large ratio of peak power to average power and increases the energy deposited per unit accelerated mass. In this way, GFPPT operation provides access to high peak exhaust velocity and specific impulse without the high average power or sustained thermal loading associated with a long, high-current discharge.

Short, high-current operation can also change how the input energy is partitioned within the plasma. Depositing greater instantaneous powers produces higher plasma temperatures and a different balance between directed acceleration and plasma heating. The local rates of mechanical work performed by the Lorentz force and resistive Joule heating are represented by $\mathbf{V}\cdot(\mathbf{J}\times\mathbf{B})$ and $\eta_eJ^2$, respectively, where $\eta_e$ is the plasma electrical resistivity. Higher-current operation increases plasma heating and electron temperature, which can reduce the resistivity and increase the plasma conductivity. \cite{Liu2024EffectsMode,Cohen1950TheGas} Although the absolute energy deposited through Joule heating may still increase, the higher current directly increases both $\mathbf{J}$ and the self-induced magnetic field $\mathbf{B}$, while the resulting stronger electromagnetic acceleration increases $\mathbf{V}$. These changes can cause the efficiency by which electrical energy is converted into kinetic energy to increase. This more favorable energy partitioning provides an additional mechanism through which the GFPPT regime can produce higher peak exhaust velocity and operate at higher efficiency.

To examine how peak current influences exhaust velocity and electromagnetic-power partitioning, a three-temperature, resistive magnetohydrodynamic model was evaluated using a series of prescribed current pulses. The duration of each pulse was adjusted with peak current to maintain an approximately constant $\int I^2(t)\,dt$, and to isolate the effect of concentrating the electromagnetic forcing into shorter, higher-current pulses. \cite{Ziemer2001ScalingThrusters} The two-dimensional axisymmetric domain represented a 5 centimeter diameter, 15 centimeter long coaxial acceleration channel followed by a plume region extending approximately 18 centimeters downstream from the accelerator exit plane. A thermal air plasma entered the channel at 110 Pa and 10,500 K, and its expansion into the initially evacuated plume was resolved using a plasma-vacuum interface-tracking scheme. \cite{Subramaniam2018AAccelerators} The finite-rate air-chemistry mechanism transported nine species, $\mathrm{N_2}$, $\mathrm{O_2}$, $\mathrm{N}$, $\mathrm{O}$, $\mathrm{N_2^+}$, $\mathrm{O_2^+}$, $\mathrm{N^+}$, $\mathrm{O^+}$, and $\mathrm{e^-}$, and included 18 reactions: four electron-impact ionization reactions, two electron-impact dissociation reactions, eight heavy-particle-impact dissociation reactions, two dissociative recombination reactions, and two electron-impact vibrational-excitation reactions. \cite{Baghirzade2026ComputationalSatellites} Varying forms of this numerical framework have previously been applied to plasma-vacuum expansion and the development of deflagration jets in coaxial plasma accelerators. \cite{Subramaniam2018ComputationalAccelerators,Baghirzade2026MHDVLEO} The complete numerical formulation, chemistry model, computational mesh details, and boundary conditions are provided in Appendix A.

\begin{figure}[tbp!]
\centering
\includegraphics[width=\figurewidth]{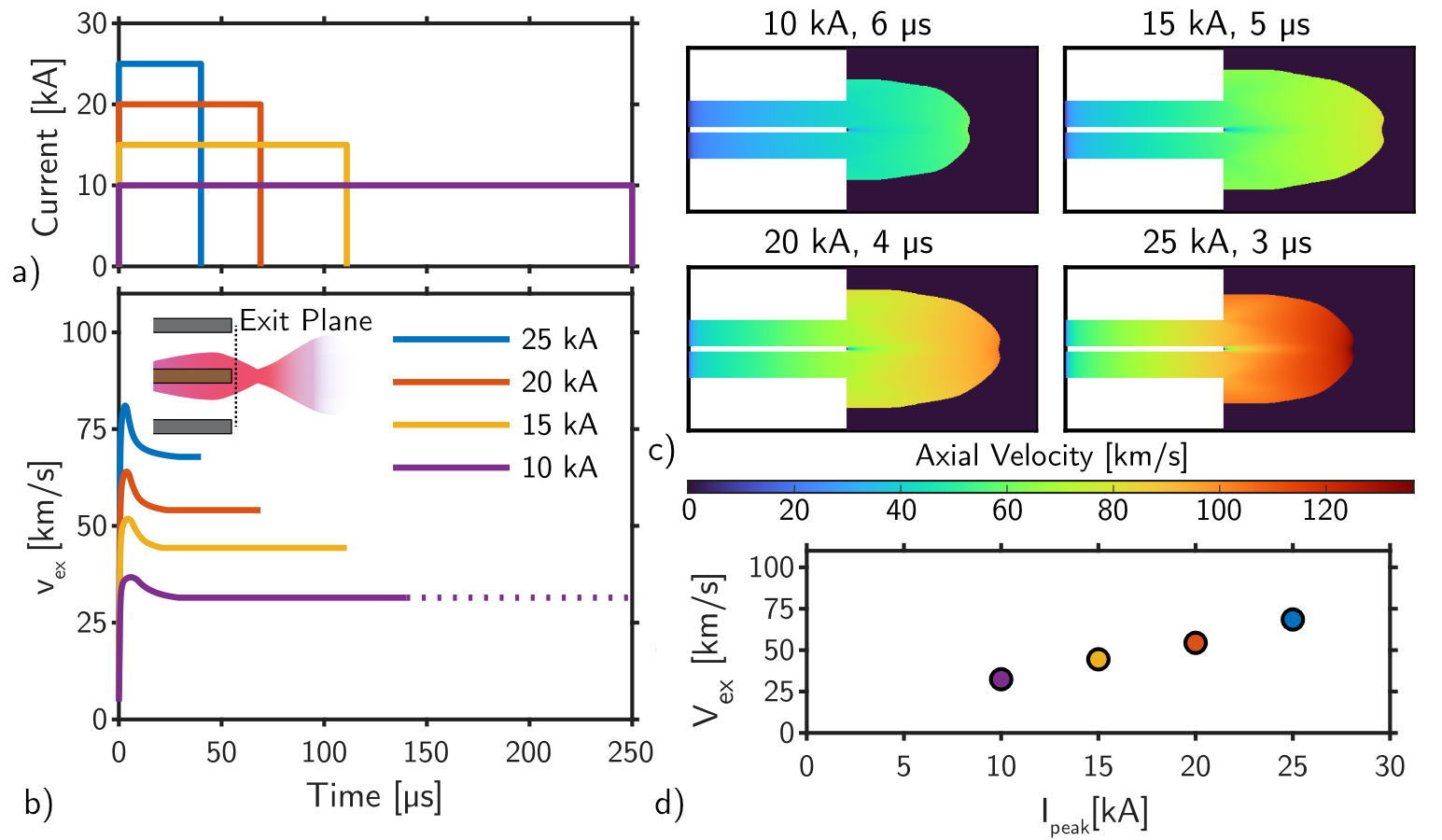}
\caption{Simulation results for the gas-fed pulsed plasma accelerator at varied peak current using a resistive MHD framework. a) Imposed current profiles for the 10, 15, 20, and 25 kA simulations. b) Temporal evolution of the mass-flux-weighted average exhaust velocity at the accelerator exit plane. c) Axial-velocity contours for the selected operating currents. d) Simulated peak exit velocity as a function of peak current.}
\label{fig:velocity-scaling}
\end{figure}

Square current pulses of 10, 15, 20, and 25 kA were prescribed with durations of 250, 110, 70, and 40~$\mu$s, respectively (Fig.~\ref{fig:velocity-scaling}a). These pulse widths, $\tau$, were selected to maintain an approximately constant value of $I^2\tau$, allowing the effects of concentrating the prescribed electromagnetic forcing into progressively shorter, higher-current pulses to be examined. \cite{Ziemer2001ScalingThrusters} The current was imposed through the corresponding azimuthal magnetic field obtained from Amp\`ere's law. The 15, 20, and 25 kA cases were simulated over their complete prescribed pulse durations. Because simulating the full 250~$\mu$s pulse was computationally prohibitive, the 10 kA case was simulated until its exhaust velocity approached a quasi-steady value, which was then extrapolated over the remainder of the prescribed pulse.

The simulations showed that increasing peak current produced stronger acceleration throughout the channel (Fig.~\ref{fig:velocity-scaling}b,c). The exhaust velocity was calculated as the mass-flux-weighted axial velocity across the accelerator exit plane, $V_{\mathrm{ex}}=\int_A\rho V_z^2\,dA/\int_A\rho V_z\,dA$ (Eq.~\ref{eq:mass_weighted_velocity}). The calculated exhaust velocity increased from 36~km/s at 10~kA to 51~km/s at 15~kA, 64~km/s at 20~kA, and 81~km/s at 25~kA (Fig.~\ref{fig:velocity-scaling}d). The local jet-tip velocity similarly increased from 61~km/s at 10~kA to 130~km/s at 25~kA. The higher-current cases also exhibited more favorable electromagnetic-power partitioning. As peak current increased, the simulated plasma resistivity decreased and the ratio of Lorentz mechanical work to Joule heating increased from approximately 1.5 at 10~kA to more than 4 at 25~kA. Concentrating the prescribed current-squared integral into shorter, higher-current pulses consequently increased both the bulk and leading-edge exhaust velocities while directing a larger fraction of the deposited electromagnetic power toward plasma acceleration rather than Joule heating. 

The high-exhaust-velocity advantage of the GFPPT regime arises from both the concentration of energy into a smaller accelerated mass and the current-dependent partitioning of electromagnetic power. At higher peak current, a greater fraction of the electromagnetic power was transferred through Lorentz mechanical work relative to Joule heating. This more favorable energy partitioning supplements the increase in energy available per unit accelerated mass. The GFPPT regime thus provides two complementary mechanisms through which short, high-current operation can increase exhaust velocity and specific impulse.

\subsection{Challenges of the GFPPT Regime}

Despite the advantages discussed previously, shortening the energy addition timescale creates two related challenges. In the GFPPT regime, $\tau_E$ is typically tens of microseconds to maintain feasible average input powers, while $\tau_G$ extends over hundreds of microseconds to milliseconds because of limitations on the mechanical response of injection valves and gas transport. \cite{Loebner2015} The resulting condition, $\tau_E\ll\tau_G$, restricts a single discharge to the propellant present during only a small fraction of the injection event, and much of the injected mass passes through the accelerator without participating in thrust generation. Short energy addition also makes performance more sensitive to the propellant distribution at breakdown. When $\tau_E\sim\tau_C$, a substantial fraction of the discharge occurs while the initial acceleration structure is traversing the accelerator. Performance then depends strongly on the propellant distribution present at breakdown and the magnetohydrodynamic mode that develops from this initial condition. \cite{Woodley2025b,Loebner2015Branch,Loebner2016,Underwood2019b,Underwood2021} A discharge delivered at an unfavorable loading condition can sacrifice much of the exhaust-velocity advantage of the GFPPT regime. Pulsed MPDs are less sensitive to this initial condition because their energy addition timescale is much longer than the accelerator flow-through timescale, $\tau_E\gg\tau_C$. An initial magneto-detonation may still form, but it occupies only a small fraction of the complete discharge and is followed by a higher-velocity magneto-deflagration that then persists for a much longer duration.

Programmable pulse shaping addresses both limitations by making peak current independently controllable from the duration and timing of energy addition. Within this expanded design space, the overlap between gas and energy addition governs propellant utilization, while peak current and the energy addition timescale govern the magnitude and duration of the Lorentz force. Peak current also interacts with the propellant loading established by gas addition to influence plasma compression, exhaust velocity, and the magnetohydrodynamic operating mode. Energy can therefore be distributed across the evolving propellant supply through multiple short, high-current discharges during a single gas-injection event. This allows the overall energy addition timescale to be extended without sacrificing the high-current acceleration of the GFPPT regime to increase propellant utilization and $I_{\mathrm{sp}}$.

\section{Experimental Setup}
\subsection{Pulsed Plasma Accelerator}

\begin{figure}[bp!]
    \centering
    \includegraphics[width=\figurewidth]{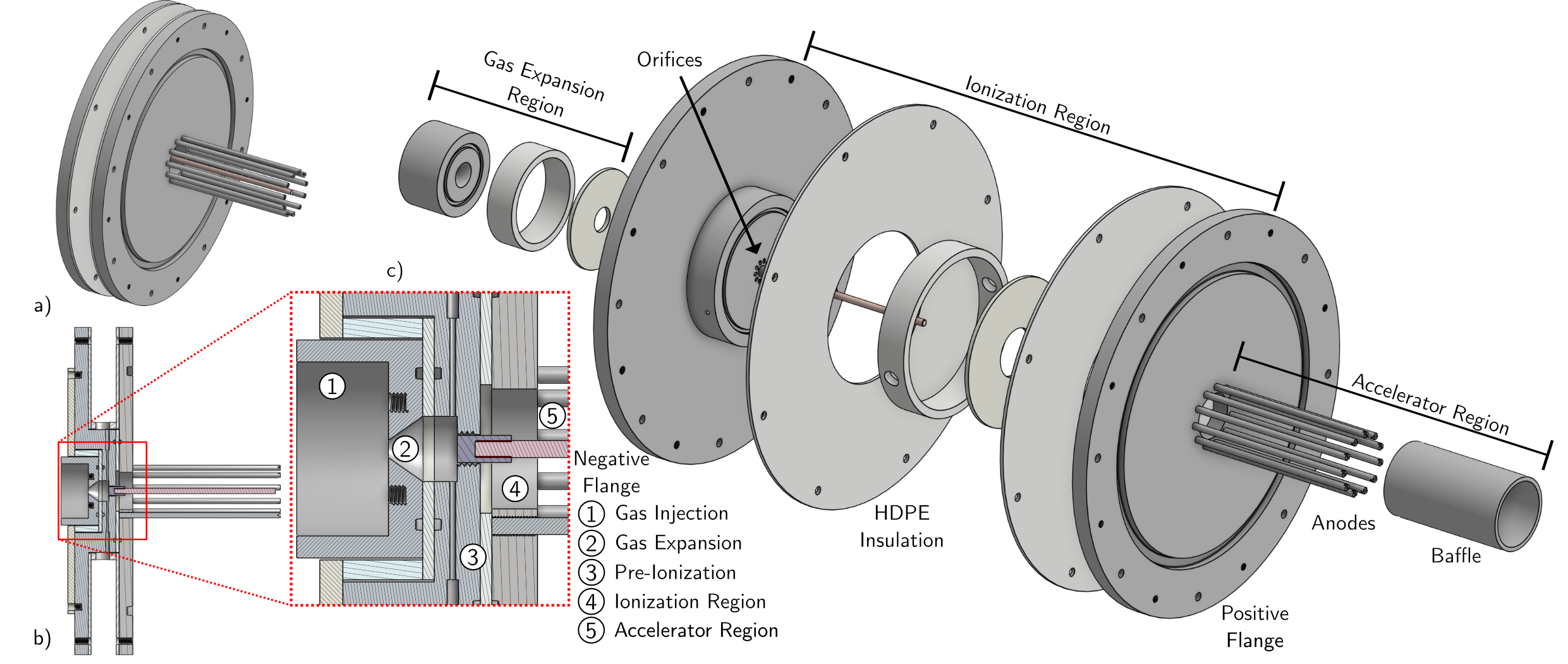}
    \caption{Pulsed electromagnetic thruster used in this study. a) Isometric view of the assembled thruster. b) Propellant path through the expansion, pre-ionization, ionization, and accelerator regions. c) Internal arrangement of the three regions.}
    \label{fig:thruster CAD}
\end{figure}

\noindent
A coaxial plasma accelerator was used to ionize and accelerate discrete injections of compressed air (Fig.~\ref{fig:thruster CAD}a). The accelerator assembly separated propellant injection, pre-ionization, propellant ionization, and electromagnetic acceleration into distinct regions (Fig.~\ref{fig:thruster CAD}b,c). This architecture allowed the injected mass bit and discharge waveform to be controlled independently, enabling systematic variation of the propellant conditions at breakdown and the subsequent acceleration process.

Compressed air first entered an expanding nozzle. The nozzle diameter increased from 5 to 19 mm over a length of 9.5 mm. The propellant then passed through a pre-ionization region containing two electrodes connected to a DC power supply. One electrode was grounded, while the other was biased to 628 V. A protection circuit controlled the pre-ionization breakdown and isolated the supply from the main thruster discharge. The circuit consisted of a 200 nF capacitor separated from the supply by a 2 k\si{\ohm} resistor and connected to the biased electrode through a 50 \si{\ohm} current-limiting resistor. A blocking diode on the grounded electrode provided additional isolation from the main discharge. The resulting seed electrons were transported with the propellant through 10 orifices, each with an area of 8.4 mm$^2$, and into the ionization region.

The ionization region was composed of two stainless steel flanges separated by high-density polyethylene and ceramic insulation. Four high-voltage leads connected to the outer edge of each flange for a total of eight high-voltage connections. Switching modules connected these leads to capacitor banks. During operation, the back flange was biased to a negative potential of up to -800 V and the front flange was biased up to 800 V. Six clusters of peaking capacitors (489.6 nF total) and bleed resistors (235 k\si{\ohm} total) were attached across the flanges. The peaking capacitors caused the voltage difference across the flanges to ring up to more than twice the capacitor bank voltage differential. This boosted voltage difference aided in ionization. The ionized propellant then advanced through the accelerator region.

Eleven stainless-steel anodes attached to the positive front flange and a central copper cathode attached between the orifices on the negative rear flange made up the accelerator region. These anode rods were used to allow visualization of the accelerator region and ensure breakdown was occurring near the gas injection plane. The anodes were 5 mm in diameter and 150 mm in length. The anodes featured a notch on the end that measured 1.5 mm across and 2 mm deep. The cathode was also 5 mm in diameter and 150 mm in length. It was secured into a stainless-steel sleeve with a threaded exterior that was screwed into the negative back flange in the center of the orifices. An aluminum baffle slid around the anodes and enclosed the accelerator region to mitigate gas leakage. The resulting plasma was then accelerated through the region and into a test section that was maintained at approximately 10$^{-7}$ Torr between discharge events using a cryogenic pump.

\subsection{Solid-State Switching Modules}

\begin{figure}[bp!]
    \centering
    \includegraphics[width=\figurewidth]{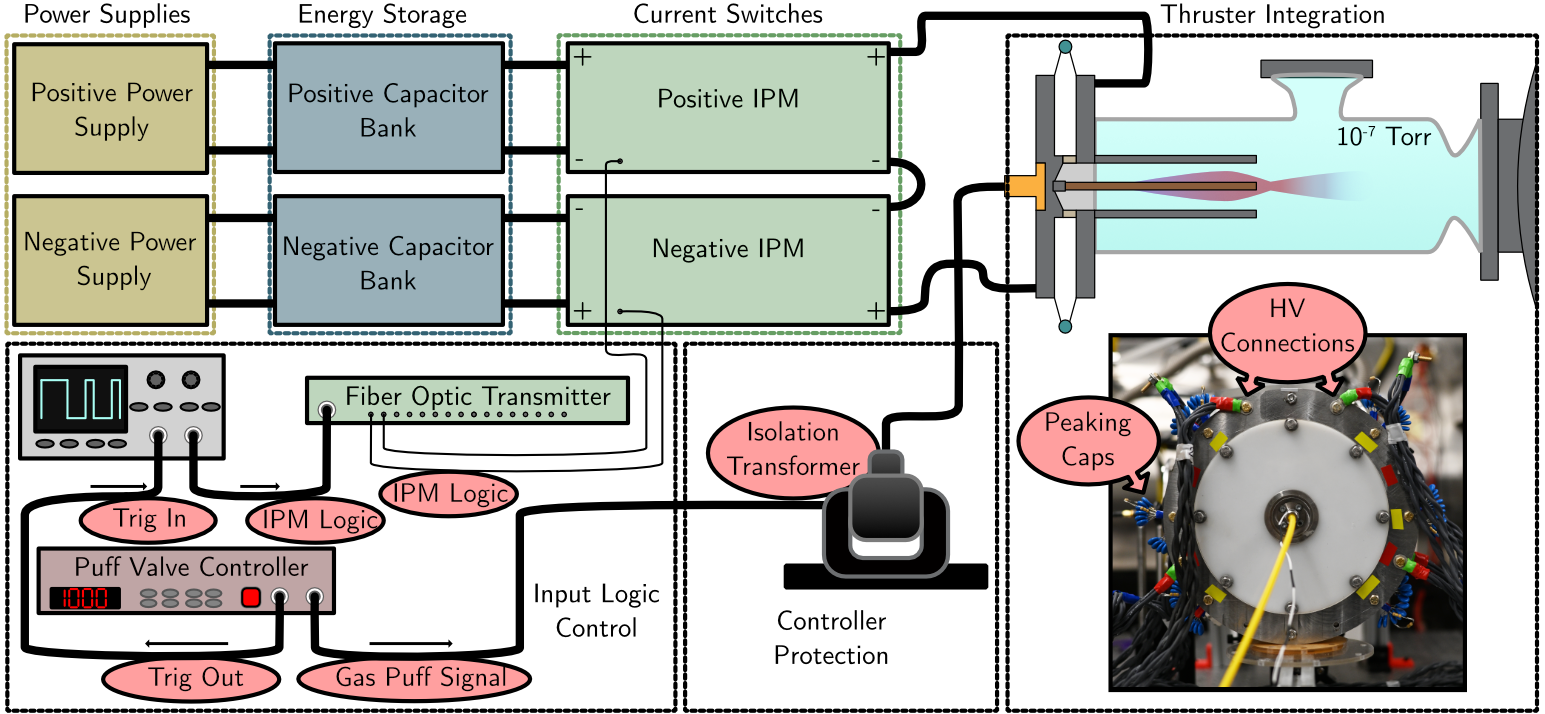}
    \caption{pulse shaping system composed of four positive-negative IPM pairs connected to the thruster. A programmable signal generator, triggered by the gas-valve controller, transmitted synchronized switching commands to the eight IPMs through fiber-optic connections. Separate high-voltage supplies charged the positive and negative module banks.}
    \label{fig:IPMs}
\end{figure}

\noindent
A system of solid-state high-current switches was used to control the current waveform in the pulsed plasma accelerator. Insulated-gate bipolar transistors (IGBTs), high-current, high-voltage semiconductor switches, can be combined in series and parallel to increase the voltage and current capability beyond that of an individual device. \cite{Bortis2008ActiveModulators,Matallana2016AnalysisFundamentals,Zorngiebel2011ModularApplications} However, differences in device characteristics, gate-driver propagation delays, and circuit parasitics can cause unequal current sharing and transient voltage imbalance. Synchronized gate drives and symmetric, low-inductance current paths are required for reliable operation of multiple switches. Parasitic inductance must also be minimized because it can increase switching losses, voltage overshoot, and oscillation during rapid switching. \cite{Anthon2014SwitchingPackage} Together with local energy storage capacitors, freewheeling diodes, and coordinated control electronics, kiloamp-level pulses can be generated with adjustable timing, duration, and repetition pattern.

In our experiments, 256 IGBTs were contained in eight integrated power modules (IPMs; Eagle Harbor Technologies) that supplied the energy required to initiate and sustain the pulsed plasma discharges. \cite{Ziemba2011EHTIGBT,Miller2013EHTIPM} Each IPM contained a parallel array of solid-state switches and $51~\mu\mathrm{F}$ of internal capacitance (Fig.~\ref{fig:IPMs}). The modules were arranged as four identical positive-negative pairs. Within each pair, one IPM was charged positively and connected to the anode, while the other was charged negatively and connected to the cathode. Simultaneously switching both modules applied approximately twice the individual module charging voltage across the thruster electrodes. The four pairs were connected in parallel at the thruster to increase the available discharge current. Dedicated twisted-pair transmission lines connected each module to the thruster, with the hot outputs connected to the corresponding electrodes and the ground outputs connected between the positive and negative modules in each pair. Each transmission line had an inductance of approximately $187.5~\mathrm{nH}$. Two independent power supplies (Magna-Power SL1000-1.5/UI, 0--1000 V, 1.5 kW) charged the positive and negative module banks to their prescribed operating voltages.

We used an in-house software package to define the IPM gate signal. This allowed us to set parameters such as peak current, pulse width, pulse shape, pulse count, interpulse spacing, and the delay between the valve opening and discharge initiation. This sequence was transferred to a programmable signal generator (Siglent SDG1062x) that was optically isolated from the IPMs by a fiber optic transmitter that delivered the switching logic to the IPM pairs. The fiber optic cable lengths were matched to within three inches of each other to minimize timing discrepancies between the IPMs. The IPMs then switched open and closed accordingly and sent current from the capacitors to the thruster.

These modules provided temporal control over the current waveform through two operating modes. Closing-switch operation resulted in the IPMs releasing all energy stored in their internal capacitor banks (51 $\mu$F per IPM). This was done by providing a signal to close the IPMs for 1 ms. In this mode, the IPMs could operate at ±800 V and up to 4 kA for a total discharge current of 16 kA and internal energy storage of 130 J. This mode provided less control over discharge parameters but still allowed for setting a delay in the discharge time relative to the beginning of gas addition. The pulse width and peak current were dictated by cable inductance with pulses lasting approximately 30--50 $\mu$s.

PWM mode provided control over the shape, duration, and duty cycle of the discharge current. In this configuration, the modules operated at up to $\pm600$ V and delivered a total current of up to 8 kA, with discharge energies exceeding 400 J. The IPMs could be actuated a maximum of 200 times within any one-second interval, with successive actuations occurring at frequencies up to 200 kHz. External capacitors containing $7200~\mu\mathrm{F}$ per IPM increased the total energy storage to more than 10 kJ and limited voltage sag during repetitive switching.

\subsection{Propellant Injection}
\noindent
A diamagnetic fast-acting gas puff valve (Parker Solenoid Valve 009-0181-900) was used to introduce a controllable amount of propellant mass into the accelerator. The valve featured a programmable opening time and the ability to adjust the gas plenum pressure. A separate IOTA power supply was used to actuate the puff valve. The actuation pulse from the power supply was used to trigger the IPM firing sequence. An isolation transformer (Stangenes SI-11422/8674962) was used to ensure the IOTA power supply was isolated from the negative back flange when the switches became active.  

To ensure that the isolation transformer did not compromise the performance of the valve, mass bit calibrations were performed at each injection condition used.  Tests were conducted at plenum pressures of 25 and 50 psig at injection durations of 600 and 1000 $\mu$s, respectively. The injections were cycled at 3 Hz over a 300 s period for a total of 900 injections for each test case. The propellant was injected into an approximately 24 L glass expansion tube that was isolated from cryogenic pumping with a gate valve. The leak rate of the container was measured, and it was subtracted from the pressure rise over the cycling duration. This overall change in pressure was used to calculate the resulting mass bit per gas injection at each condition. The mass bit was 12.86 $\mu$g for the 25 psig, 600 $\mu$s case, and it was 31.52 $\mu$g for the 50 psig, 1000 $\mu$s case. The 12.86 $\mu$g conditions were used for the $I_{\mathrm{sp}}$ study, and the 31.52 $\mu$g case was used for all other studies.

\subsection{Plasma Flow Visualization}

\noindent
Videos of the broadband plasma self-emission were acquired to visualize plume evolution during each discharge. This was used to identify exhaust structures present with different thruster operating conditions and to measure the time-of-flight velocity of jets as they expanded into vacuum. A Shimadzu HPV-X2 high-speed camera was operated at frame rates between 500 thousand and 10 million frames per second, with exposure times ranging from 100 to 2000 ns. The camera’s resolution was 50,000 or 100,000 pixels, depending on the frame rate. \cite{Underwood2019a} The system featured a circular buffer of 256 frames, which was used to capture each discharge event. A 50 mm lens (Nikon AF-S Nikkor 50 mm 1:1.8G) was used to capture side-on images. Neutral-density filters ranging from 1 to 3 were used to reduce light intensity from the plasma. The camera was focused on the thruster exit plane and was triggered externally by the IOTA output current with a specified delay. 

\subsection{Pendulum Thrust Stand}

\begin{figure}[bp!]
    \centering
    \includegraphics[width=\figurewidth]{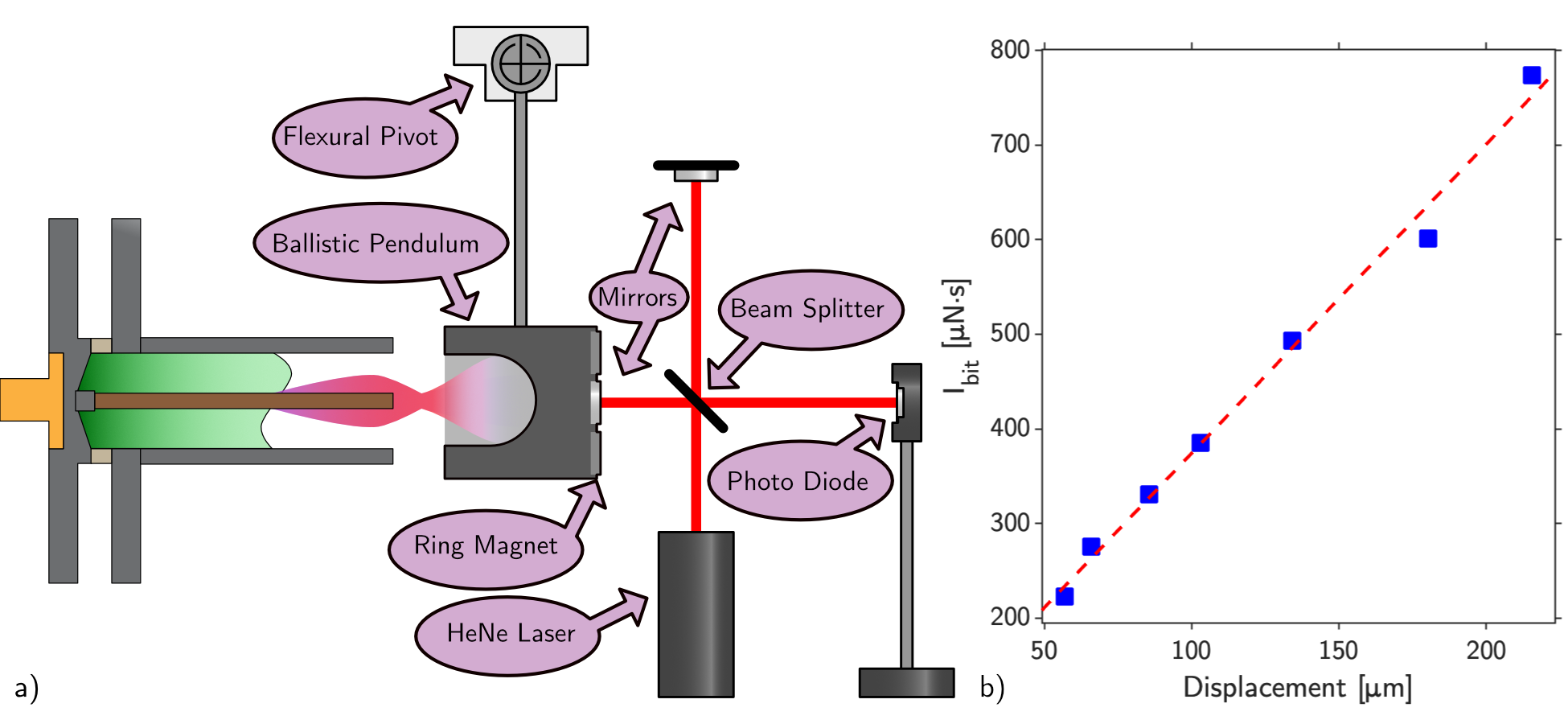}
    \caption{Ballistic pendulum and interferometer setup for this study. a) Thrust was measured using a ballistic pendulum, and its displacement was measured using a Michelson interferometer. b) A voice coil was used to calibrate the pendulum by applying known impulses to a ring magnet affixed to the back of the pendulum. The measured displacements and the known impulses were used to create a calibration curve. Calibration was performed in the test section.}
    \label{fig:pendulum}
\end{figure}

\noindent
Impulse bits from the thruster were measured using a ballistic pendulum mounted in the test section (Fig. ~\ref{fig:pendulum}a). The pendulum head was positioned 7 cm downstream of the accelerator exit plane, and the plume induced a backward deflection proportional to the delivered impulse. The pendulum body was fabricated from aluminum in a fishtrap geometry sized to be equal to the thruster volume. This geometry minimized reflection, stagnation effects, and secondary momentum losses. The total pendulum mass was 250 g, the pendulum arm length was 12 cm, and the suspension stiffness was $k$ = 21.2 N$\cdot$m/rad to maximize displacement sensitivity. 

A Michelson interferometer measured the displacement of the pendulum used to infer impulse bits. The interferometer employed a stabilized HeNe laser (coherence length $>$100 m, Thorlabs HRS015B) and a photodetector (Thorlabs SM05PD1A). A mirror mounted on the pendulum provided one arm of the interferometer, such that its motion produced measurable fringe shifts. Measurements were taken only after the pendulum returned to baseline oscillations. Insulated mounts on the vacuum chamber suppressed table vibrations. A permanent magnet was affixed to the pendulum to facilitate controlled calibration and correlate applied force with displacement. A piezoelectric crystal with a resonance frequency of 8 kHz was attached to one interferometer mirror to frequency-lock the photodetector. Fringe shifts were counted using a directional fringe-counting algorithm that separated forward and backward motion, enabling accurate determination of maximum displacement. The peak displacement was then compared against calibration data relating displacement to impulse, allowing estimation of the delivered impulse bit.

The pendulum was calibrated using a voice coil that was placed inside the vacuum test section. The coil was characterized first externally by mounting the pendulum on a precision balance, applying a known DC current, and recording the steady diamagnetic repulsive force produced by the voice coil. This procedure yielded a direct force–current relationship. The same coil was then installed in the vacuum chamber, and the duration of the applied current pulse was varied from 1 $\mu$s to over 100 $\mu$s to impart controlled impulses to the pendulum. The resulting peak displacements were recorded interferometrically and used to generate a calibration curve relating the pendulum’s displacement to delivered impulse bit (Fig.~\ref{fig:pendulum}b). The measured $I_{\mathrm{bit}}$ ranged from 222.6 $\mu$N$\cdot$s to 773.97 $\mu$N$\cdot$s. All measured thruster impulse bits were subsequently computed using this displacement-to-impulse calibration. 

\section{Results and Discussion}

\subsection{Programmable Waveform Control Using Solid-State Switching Modules}

\begin{figure}[tbp!]
\centering
\includegraphics[width=\figurewidth]{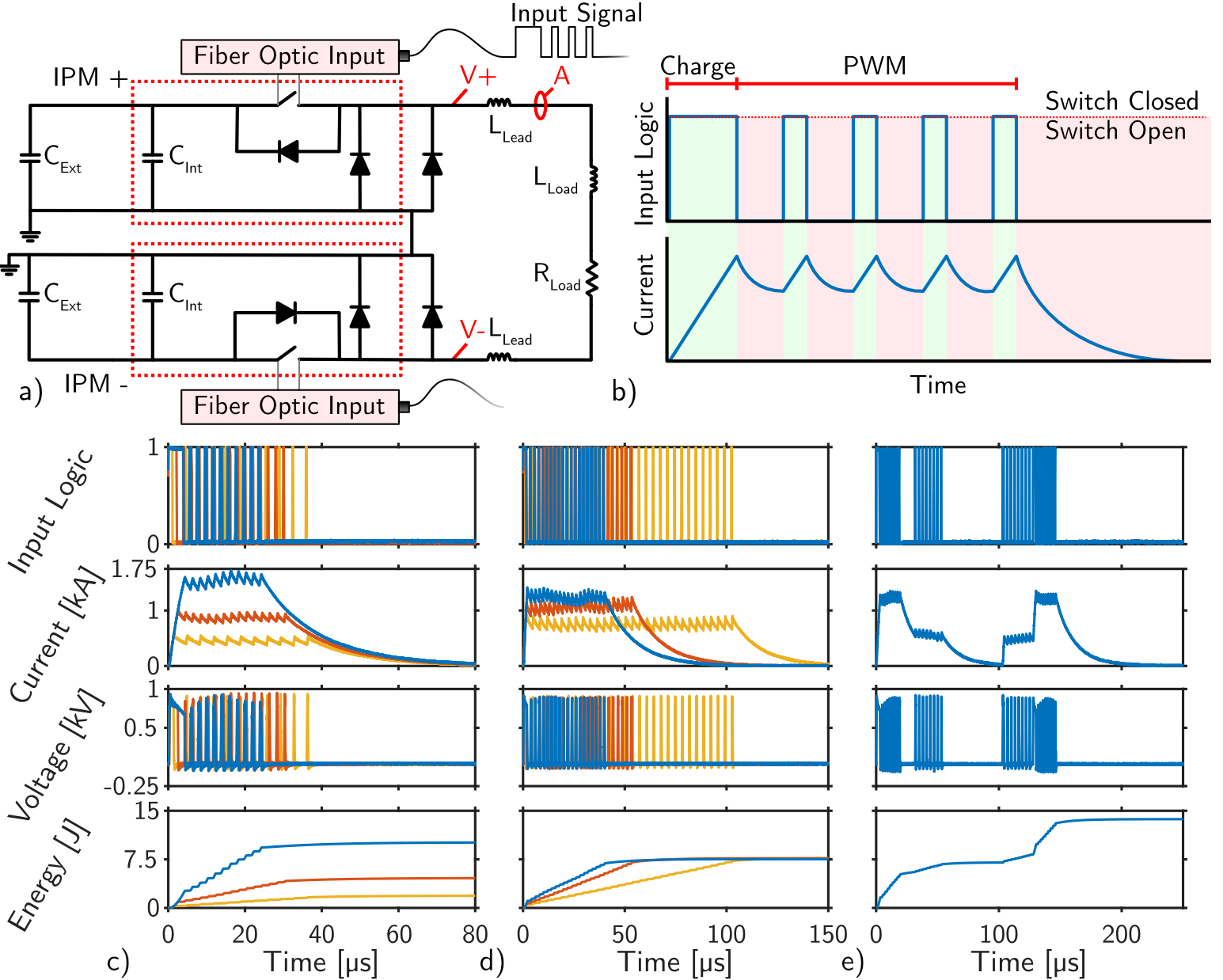}
\caption{Test-load validation of programmable IPM waveform control. a) Circuit containing one positive-negative IPM pair and a test load, with $R_{\mathrm{Load}}=100$ m\si{\ohm}, $L_{\mathrm{Load}}=1.324~\mu$H, $L_{\mathrm{Lead}}=187.5$ nH, $C_{\mathrm{Int}}=51~\mu$F, and $C_{\mathrm{Ext}}=7.2$ mF. b) Input logic and resulting load current during PWM operation. c) Energy variation at a fixed pulse width of 25 $\mu$s. d) Pulse-duration variation from 40 to 100 $\mu$s at a fixed energy of 7.5 J. e) Programmable multilevel current waveform.}
\label{fig:test_load}
\end{figure}

\noindent
Current waveform control was evaluated using a representative test load consisting of a resistor array (100 m\si{\ohm}) and an inductor (1.324 $\mu$H) chosen to simulate thruster operation while isolating IPM behavior from the varying impedance of the plasma. Both single-pulse and burst operation were tested (Fig.~\ref{fig:test_load}a). A single pair of IPM modules was used for these initial tests, and both the load voltage and load current were measured to quantify the response of the switching system. These measurements provided a direct assessment of how effectively the input logic translated into controlled discharges. 

The system was operated so that each pulse featured a charging interval and a pulse-width-modulation interval. During the charging interval, the input logic was held high, closing the switch pairs and allowing current through the load to rise. This produced a steep current ascent limited primarily by the combined inductance of the load and wiring (Fig. ~\ref{fig:test_load}b). Once the current reached the desired amplitude, the logic transitioned into the PWM interval, which regulated the current by modulating the switch state on microsecond timescales. The PWM interval began with a low command that opened the switches and prevented current from flowing through the IPMs. During this off portion, the load current decayed at a rate governed by the inductance and resistance of the circuit. Freewheel diodes placed at the IPM outputs provided a controlled current path during this segment, preventing voltage spikes and enabling a smooth decay trajectory. The logic then returned high for a short on-portion, raising the current back to its previous level. This cycle repeated until the desired total pulse duration was reached. The current was then allowed to decay.

Three key capabilities were demonstrated using a charging voltage of 400 V (800 V difference across IPMs). First, the system enabled control of energy deposition at fixed pulse width. Increasing the charging interval produced higher peak currents and larger pulse energies while maintaining identical total pulse lengths (Fig.~\ref{fig:test_load}c). The PWM input logic was unique to each case as different signals were required to maintain each current plateau. With charging voltage constant, peak currents of 0.5 kA, 1 kA, and 1.5 kA were used to achieve pulse energies of 2 J, 5 J, and 10 J, respectively. Second, pulses of different durations (measured from the initial current rise to the end of PWM) were shaped to deliver nearly identical total energies by adjusting the charging portion and PWM timing, allowing controlled comparisons across pulse lengths from tens to hundreds of microseconds (Fig.~\ref{fig:test_load}d). Three pulse widths were used: a 40 $\mu$s pulse with a current plateau of 1.25 kA, a 50 $\mu$s pulse with a current plateau of 1 kA, and a 100 $\mu$s pulse with a current plateau of 0.8 kA. These three pulses all resulted in a pulse energy of 7.5 J. Third, the modular input structure allowed arbitrary temporal shaping of the current waveform, including stepped profiles, tapered decays, and complex sequences not easily achievable with conventional pulse-forming networks (Fig.~\ref{fig:test_load}e). A complex input signal was used to demonstrate this. Current rose initially to 1.15 kA and held for 20 $\mu$s before dropping to 0.5 kA for 30 $\mu$s. This pulse was succeeded by another with the inverted behavior of the first.

\begin{figure}[bp!]
\centering
\includegraphics[width=\figurewidth]{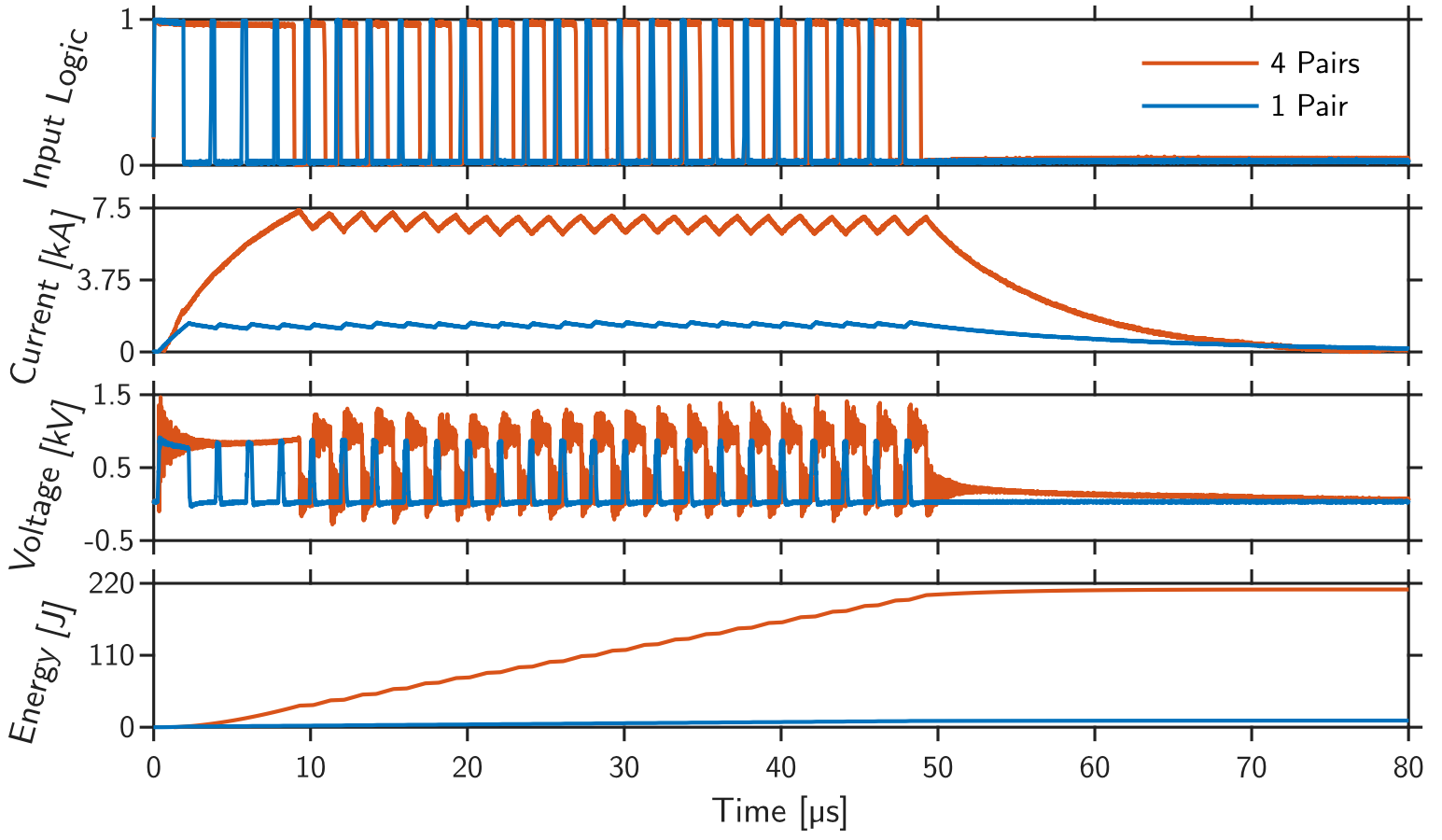}
\caption{Discharge characteristics of the full eight-switch system (four pairs) in comparison to two switches (one pair). The full system can maintain currents of 8 kA in PWM mode while a single pair is limited to 2 kA in PWM mode. The system-test voltage was measured across the load rather than directly at the IPM outputs (Fig.~\ref{fig:test_load}).}
\label{fig:system_test}
\end{figure}

Four IPM pairs were connected in parallel to scale the current and pulse energy for thruster operation (Fig.~\ref{fig:system_test}). Each pair was supplied by separate positive and negative external capacitor banks charged to 400 and $-400$ V, respectively. The full system increased the available current from 2 kA with one pair to approximately 8 kA and supported discharge energies exceeding 200 J. The additional wiring and interconnects increased the loop inductance, slowing the current rise and decay and increasing the charging interval required for a 50 $\mu$s pulse from 1 to 9 $\mu$s. Despite this change, programmable control of pulse width, energy, and waveform shape was retained at full-system scale.

\begin{figure}[tp!]
\centering
\includegraphics[width=\figurewidth]{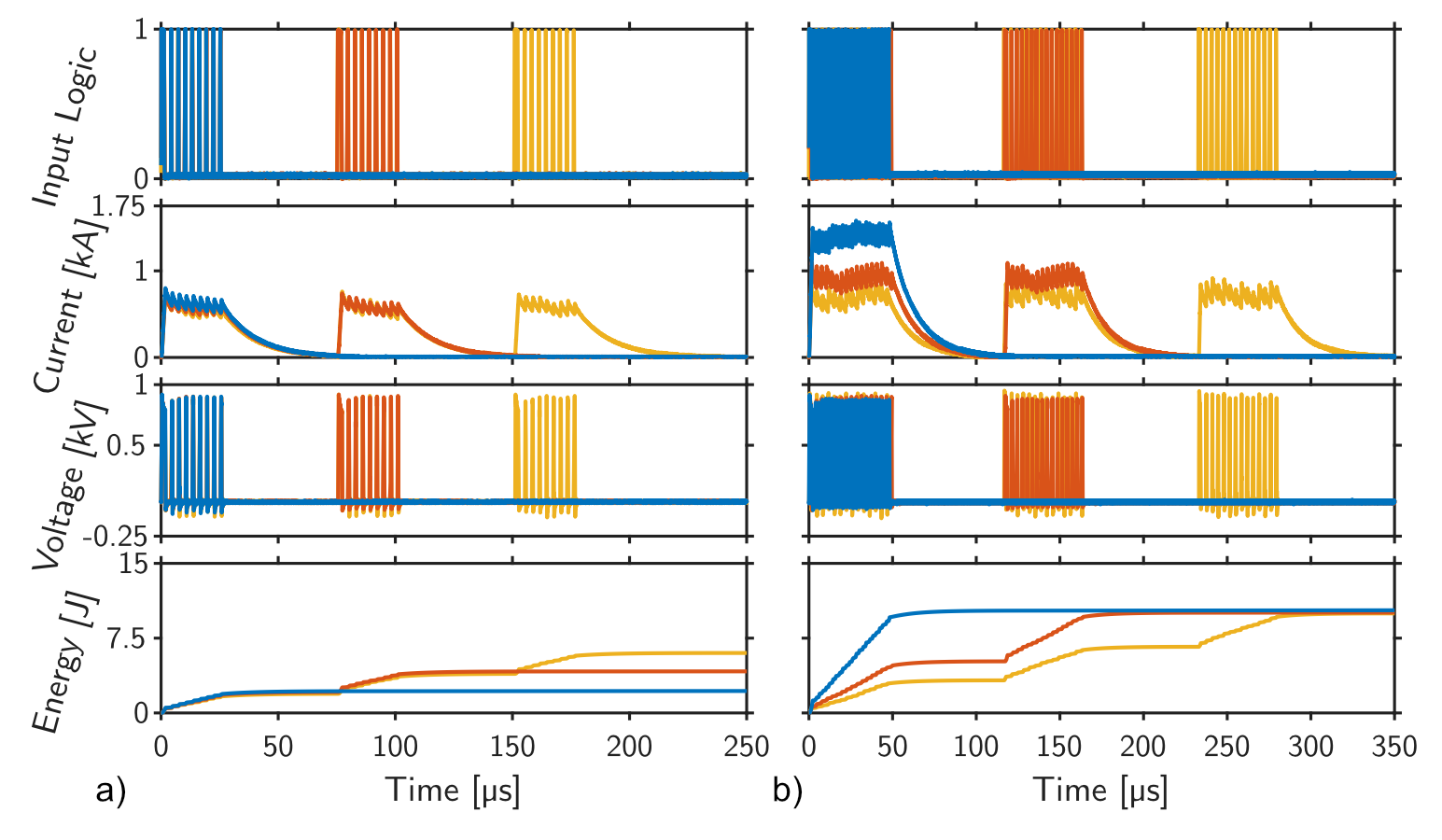}
\caption{Input signals, current and voltage traces, and discharge energies for burst firing mode. a) Bursts containing N = 1, 2, and 3 pulses of 25 $\mu$s were fired while the energy of each individual pulse remained constant.  b) N = 1, 2, and 3 pulses of 50 $\mu$s were fired holding the total energy of the burst at 10 J. }
\label{fig:burst_test_load}
\end{figure}

Burst-mode operation was evaluated using a single IPM pair connected to the same test load and charged to $\pm400$ V. Each pulse contained an initial charging interval followed by a PWM-regulated maintenance interval. Bursts containing $N=1$--3 pulses of 25 $\mu$s were first operated with a 0.5 kA current plateau, an energy of 2 J per pulse, and an interpulse gap of 50 $\mu$s, allowing the current to return nearly to zero before each subsequent pulse (Fig.~\ref{fig:burst_test_load}a). The rise time, peak current, and PWM-regulated plateau remained consistent across the burst, while the total energy increased linearly with pulse count. The charging interval and PWM switching were then adjusted to maintain a total burst energy of 10 J as the number of 50 $\mu$s pulses increased (Fig.~\ref{fig:burst_test_load}b). The corresponding current plateaus decreased from 1.4 kA at $N=1$ to 1.0 and 0.7 kA at $N=2$ and 3, respectively. These results demonstrated independent control over pulse count and total burst energy with limited capacitor-voltage sag.

\subsection{Inducing Mode Transitions Using Programmable Discharge Delays}

\begin{figure}[tbp!]
\centering
\includegraphics[width=\figurewidth]{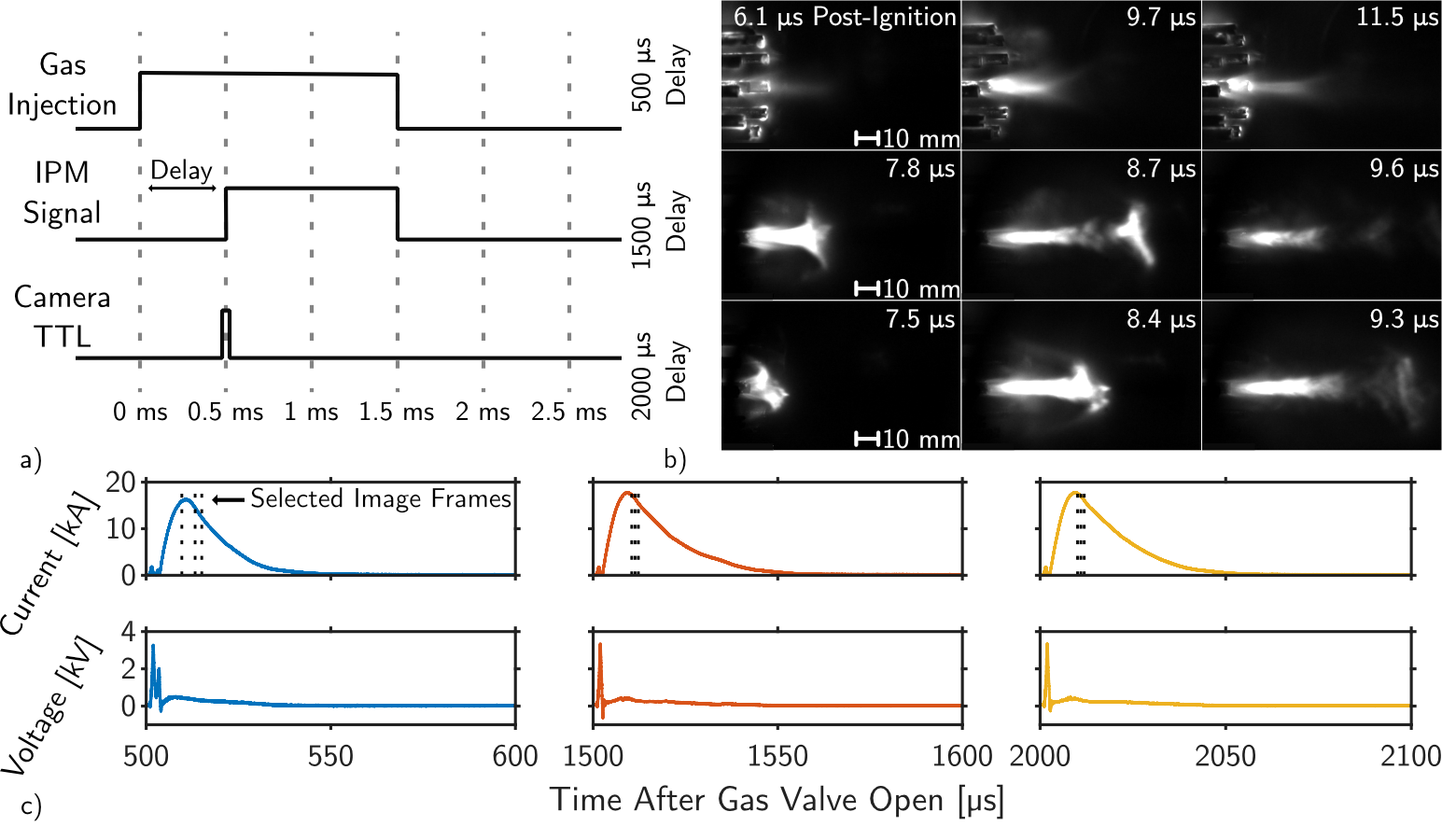}
\caption{Plume evolution was captured using high-speed imaging to understand the effect of delays on the acceleration structure. a) Timing diagram of the gas injection, IPM gate signal, and camera TTL. b) High-speed images show the differences in acceleration structure. The 500 $\mu$s delay case depicts a magneto-deflagration while the 1500 and 2000 $\mu$s delays depict a magneto-detonation. c) The current and voltage traces show less ringing from the peaking capacitors at larger delays, consistent with delay-dependent differences in propellant loading.}
\label{fig:delay}
\end{figure}

\noindent
Experiments were performed to determine whether programmable delays alone could select the acceleration mode. The gas-injection and electrical conditions were held fixed while the delay between gas-valve actuation and discharge initiation was varied. A plenum pressure of 50 psig and an opening time of 1500 $\mu$s were used, and delays of 500, 1500, and 2000 $\mu$s were imposed relative to the beginning of the gas injection to create different propellant loading conditions (Fig.~\ref{fig:delay}a). The pulsed power supply was operated in closing-switch mode with a charging voltage of $\pm$650 V, resulting in a peak current of approximately 16 kA. Side-on images were taken to classify acceleration structures, and current and voltage traces were measured to identify breakdown and circuit characteristics. 

Broadband images of the plasma self-emission showed that the plume structure changed substantially with the programmable delay. At a delay of 500 $\mu$s, the discharge produced a smooth, continuously expanding plume characteristic of magneto-deflagration-driven acceleration (Fig.~\ref{fig:delay}b). At delays of 1500 and 2000 $\mu$s, a bright, spatially concentrated current sheet formed and propagated through the accelerator. This current sheet was followed by a broader plasma expansion. The current sheet indicates that the discharge current, Joule heating, and electromagnetic acceleration were concentrated within a narrow ionization front propagating into the neutral gas accumulated during the longer delay. \cite{Underwood2021,Loebner2015Branch} Its snowplow-like structure was consistent with magneto-detonation-driven acceleration and was particularly visible after leaving the accelerator in the 1500 $\mu$s case at 8.7 $\mu$s after discharge initiation (Fig.~\ref{fig:delay}b). As the current sheet departed the accelerator, it became canted relative to the exit plane, revealing that the ionization and acceleration front did not propagate uniformly across the channel. This canting indicates an azimuthal asymmetry in the current distribution, local plasma conductivity, or upstream mass loading rather than a purely axisymmetric exhaust structure. After the current sheet exited, the remaining plasma expanded into a smoother, more diffuse structure characteristic of magneto-deflagration. These observations show that changing the delay between gas addition and energy deposition controlled not only the initial mass distribution within the accelerator, but also whether acceleration began as a distributed deflagration or as a localized, nonuniform magneto-detonation current sheet.

The ability to select the initial acceleration mode through discharge timing adds propellant distribution as a controllable dimension of thruster operation without requiring changes to the accelerator geometry or electrical pulse. This distinction is especially important for short-pulse accelerators, in which the initial acceleration structure persists through a substantial fraction of the energy addition interval. Whether the discharge develops as a distributed magneto-deflagration or a localized magneto-detonation can influence current localization, plasma heating, electromagnetic momentum coupling, and electrode loading. Coordinating discharge initiation with the evolving propellant distribution provides a means of tailoring both the acceleration process and the conditions under which electrical energy is coupled to the plasma.

\subsection{Performance Scaling Across Operating Regimes}

\begin{figure}[tbp!]
\centering
\includegraphics[width=\figurewidth]{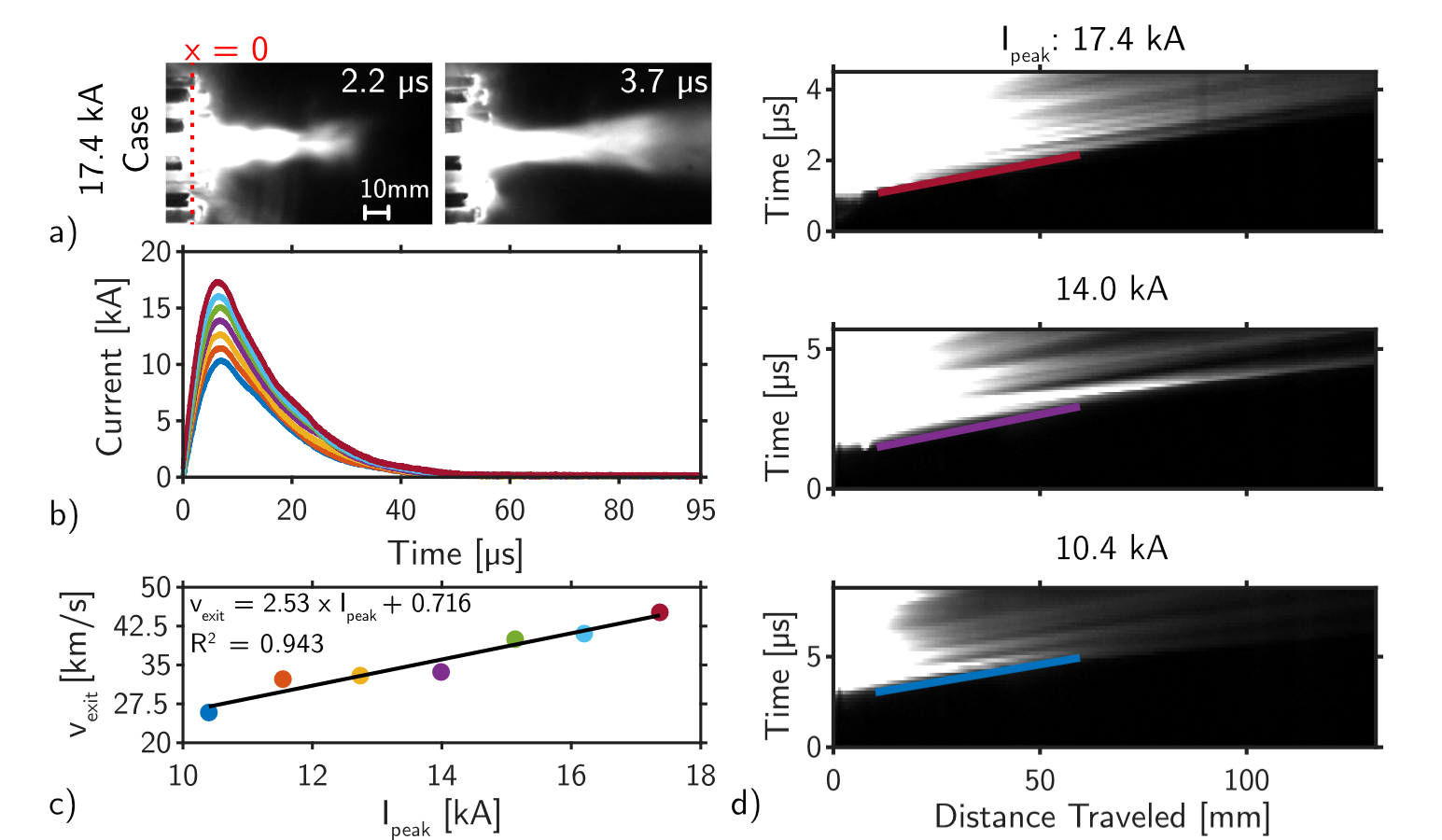}
\caption{Time-of-flight measurements used to characterize the relationship between plume leading-edge velocity and peak current during short-pulse, closing-switch operation. 
a) Representative high-speed images showing plume propagation from the accelerator exit plane for the 17.4 kA case. 
b) Current traces for seven operating conditions with peak currents ranging from 10.4 to 17.4 kA. 
c) Plume leading-edge velocity as a function of peak current, showing an approximately linear relationship. 
d) Streak maps and fitted leading-edge trajectories for the 10.4, 14.0, and 17.4 kA cases.}
\label{fig:PWTOF}
\end{figure}

\noindent
Time-of-flight measurements were conducted to characterize how plume-front velocity varied with peak current (Fig.~\ref{fig:PWTOF}). After a discharge delay of 1000 $\mu$s, air was injected from the plenum at 50 psig with a gas-injection time of 1500 $\mu$s. Charging voltages from 400 to 700 V produced peak currents ranging from 10.4 to 17.4 kA, while the pulse width remained approximately constant at 20-30 $\mu$s (Fig.~\ref{fig:PWTOF}b). High-speed images captured the plume propagation from the accelerator exit plane, and streak maps were constructed to determine the leading-edge trajectory for each condition (Figs.~\ref{fig:PWTOF}a and \ref{fig:PWTOF}d).

The measured leading-edge velocity obtained through time-of-flight measurements increased approximately linearly with peak current. The velocity rose from approximately 24 km/s at 10.4 kA to 45 km/s at 17.4 kA (Fig.~\ref{fig:PWTOF}c). The magnitudes of these time-of-flight velocities are lower than the simulated velocities (Fig.~\ref{fig:velocity-scaling}d), but the linear scaling is consistent. Although the leading-edge velocity does not represent the mass-flux-weighted effective exhaust velocity directly, the trend demonstrates that higher-current, short-duration discharges produced faster plume propagation and is consistent with stronger electromagnetic acceleration. This result establishes the physical basis for the subsequent thrust measurements, which determine whether the velocity advantage of high-current, GFPPT-like pulses translates into increased impulse bit. 

To determine how this increase in peak current translated into delivered impulse and thrust-to-power ratio, peak current was varied while pulse width was held approximately constant within two sets of experiments. Because increasing current at fixed pulse width increased the deposited energy, these tests characterized the performance scaling with peak current and pulse energy at a nearly fixed energy addition timescale. The thruster was operated at a plenum pressure of 50 psig, a gas-injection duration of 1000 $\mu$s, and a discharge delay of 800 $\mu$s. Four closing-switch discharges lasting approximately 20-30 $\mu$s produced peak currents of 10, 12, 13, and 16 kA and discharge energies of 31, 46, 57, and 76 J, respectively. Three additional 50 $\mu$s PWM discharges maintained current plateaus of 4, 6, and 8 kA and deposited 17, 38, and 77 J, respectively. Axial high-speed images were acquired for the 4 and 8 kA PWM cases to examine changes in plume structure with peak current. Current through the accelerator and voltage across the positive and negative flanges were measured to determine the electrical waveforms and deposited energy for each discharge.

\begin{figure}[tbp!]
\centering
\includegraphics[width=\figurewidth]{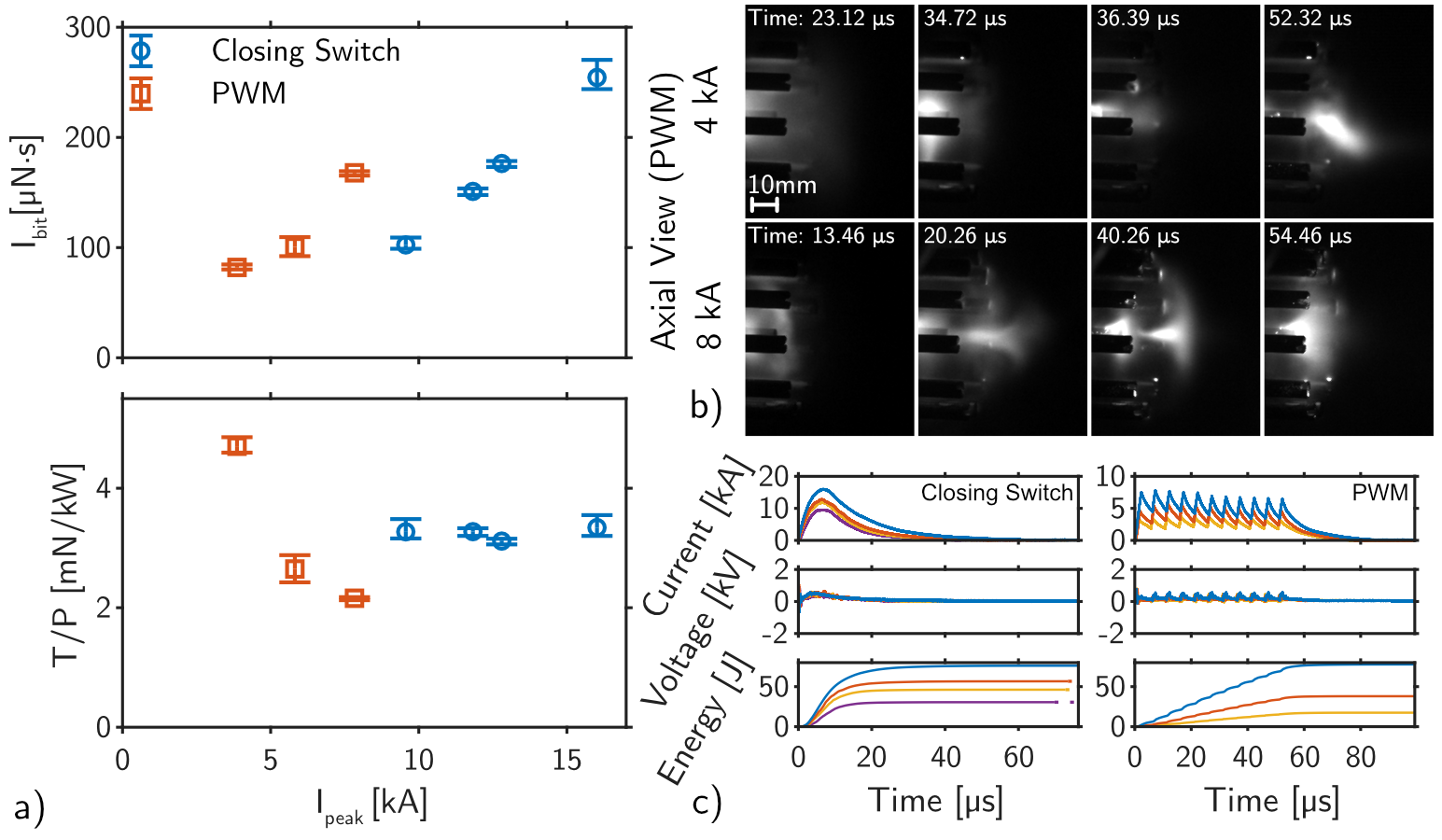}
\caption{Effect of peak current and discharge energy at approximately fixed pulse width. In closing-switch mode, the thruster was operated at peak currents from 10 to 16 kA with discharge durations of approximately 20-30 $\mu$s. In PWM mode, current plateaus from 4 to 8 kA were maintained for 50 $\mu$s. 
a) $I_{\mathrm{bit}}$ increased with peak current and discharge energy, while $T/P$ remained approximately constant within each operating series. 
b) Axial high-speed images of 4 and 8 kA PWM cases. 
c) Current, voltage, and cumulative energy traces showing the distinct energy addition profiles produced by closing-switch and PWM operation.}
\label{fig:pulse_energy}
\end{figure}

For both closing-switch and PWM operation, $I_{\mathrm{bit}}$ increased with peak current and discharge energy, while $T/P$ remained approximately constant within the experimental scatter (Fig.~\ref{fig:pulse_energy}a). Because $T/P=I_{\mathrm{bit}}/E$, this approximately constant ratio indicates that the impulse bit increased nearly proportionally with deposited energy in both operating modes. This behavior is consistent with self-field electromagnetic scaling because both the electromagnetic impulse and deposited energy depend approximately $\int I^2(t)\,dt$. By increasing $I_{\mathrm{bit}}$ while holding $T/P$ nearly constant for fixed mass bits, thrust efficiency was increased from 0.5\% to 1.5\% as current increased from 10 to 16 kA.

The high-speed images showed a clear change in plume structure with current (Fig.~\ref{fig:pulse_energy}b). The 4 kA case produced a broad, asymmetric plume with weak radial compression and visible magnetohydrodynamic instabilities. \cite{Underwood2019c} In contrast, the 8 kA case developed a narrower and more axially directed plume, with luminous exhaust extending farther downstream at equivalent times after discharge initiation. Although some asymmetries remained at 8 kA, the stronger radial compression was consistent with the increase in the self-field Lorentz force expected at higher current. The more collimated plume structure was also consistent with the larger measured impulse bit, suggesting that a greater fraction of the accelerated momentum was directed axially.

\begin{figure}[bp!]
\centering
\includegraphics[width=\figurewidth]{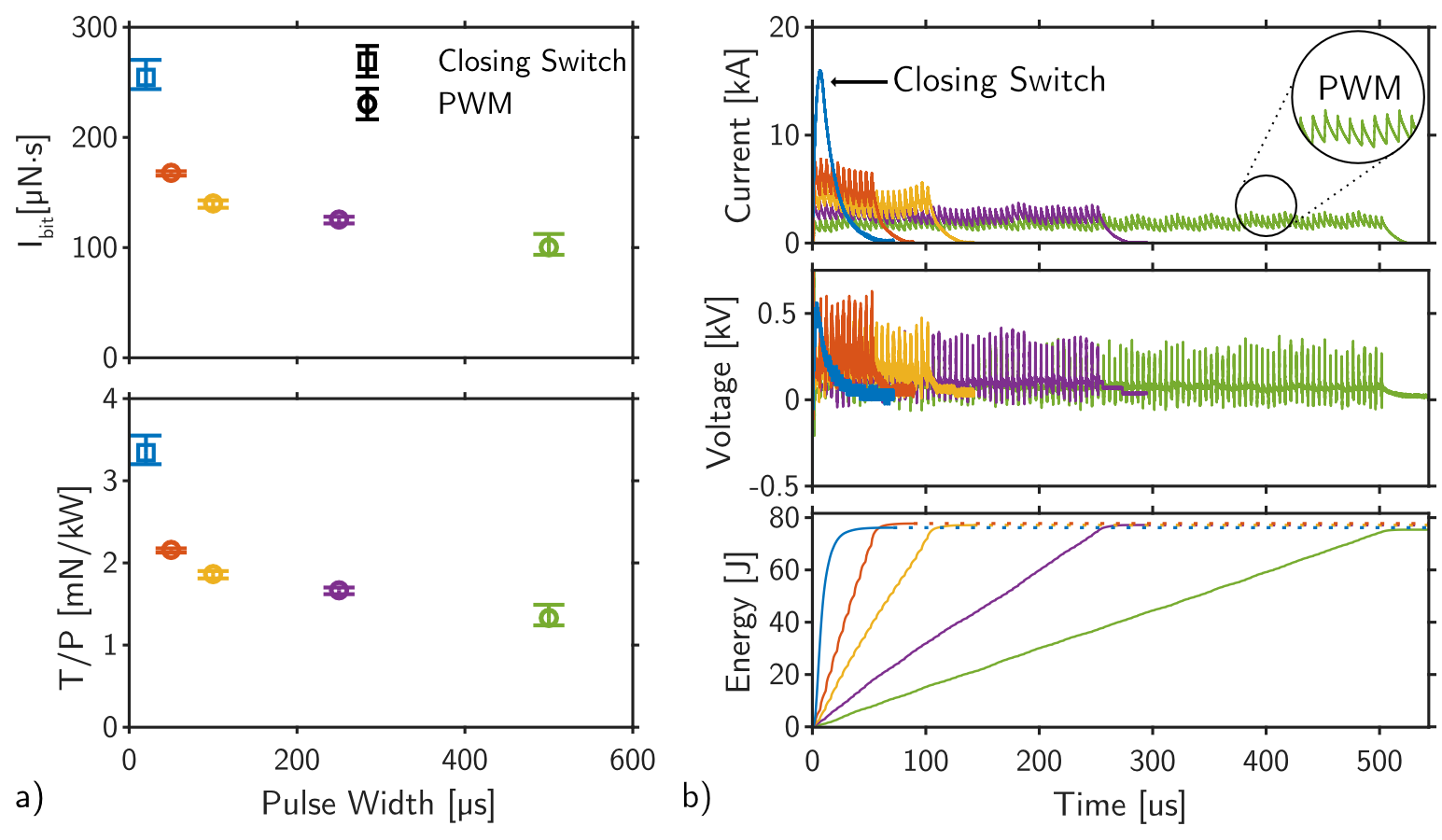}
\caption{Effect of pulse duration and operating current at nearly fixed discharge energy and injected mass bit. Five pulses ranging from approximately 30 to 500 $\mu$s deposited 75--78 J, while peak current decreased from 16 to approximately 2 kA. a) $I_{\mathrm{bit}}$ and $T/P$ decreased as the energy addition timescale increased. b) Current, voltage, and cumulative energy traces. Deposited energies were matched within 4\%, and the PWM cases exhibited switching-induced current fluctuations that increased in amplitude with plateau current.}
\label{fig:pulse_width}
\end{figure}

To directly compare short, high-current GFPPT-like operation with longer, lower-current pulsed-MPD-like operation, the discharge energy and nominal injected mass bit were held approximately constant while the pulse duration was varied. The thruster was operated at a plenum pressure of 50 psig, a gas-injection duration of 1000 $\mu$s, and a discharge delay of 800 $\mu$s. Five single-shot discharges were tested with energies between 75 and 78 J. The shortest case was a 20-30 $\mu$s closing-switch discharge with a peak current of 16 kA and an energy of 76 J. The remaining cases were produced using PWM: a 50 $\mu$s pulse with a 6 kA current plateau and 78 J discharge energy, a 100 $\mu$s pulse with a 4 kA plateau and 75 J discharge energy, a 250 $\mu$s pulse with a 3 kA plateau and 75 J discharge energy, and a 500 $\mu$s pulse with an approximately 2 kA plateau and 75 J discharge energy. Current through the accelerator and voltage across the positive and negative flanges were measured to characterize the electrical waveforms and verify the deposited energy for each condition. The corresponding impulse bits were measured to determine how redistributing a nearly fixed amount of energy over different pulse durations affected the conversion of electrical energy into axial impulse.

At nearly fixed discharge energy and fixed injected mass bit, the delivered impulse decreased substantially as the pulse width increased and the operating current decreased. For pulses depositing between 75 and 78 J, $I_{\mathrm{bit}}$ decreased from approximately 250 $\mu$N$\cdot$s for the 20-30 $\mu$s closing-switch pulse to approximately 95 $\mu$N$\cdot$s for the 500 $\mu$s PWM pulse (Fig.~\ref{fig:pulse_width}a). The longer, lower-current, pulsed-MPD-like cases produced less impulse from approximately the same deposited energy than the shorter, higher-current, GFPPT-like cases. In contrast to the fixed-pulse-width study, $T/P$ decreased as the operating condition shifted toward longer pulse durations and lower currents. We speculate this is because of collimation of the plume or changes in how energy is partitioned into thermal heating due to changes in ionization and electrical conductivity at differing levels of current.

These results show that extending the discharge to interact with more of the injected propellant does not necessarily improve performance. At lower current, additional propellant may be ionized or heated without being converted efficiently into directed axial momentum. For the conditions examined here, concentrating the available energy into a short, high-current pulse produced substantially more impulse, motivating the use of timing or multiple short pulses to improve propellant utilization while preserving strong electromagnetic acceleration.

\subsection{Extension to Micro-Bursts}
\begin{figure}[bp!]
\centering
\includegraphics[width=\figurewidth]{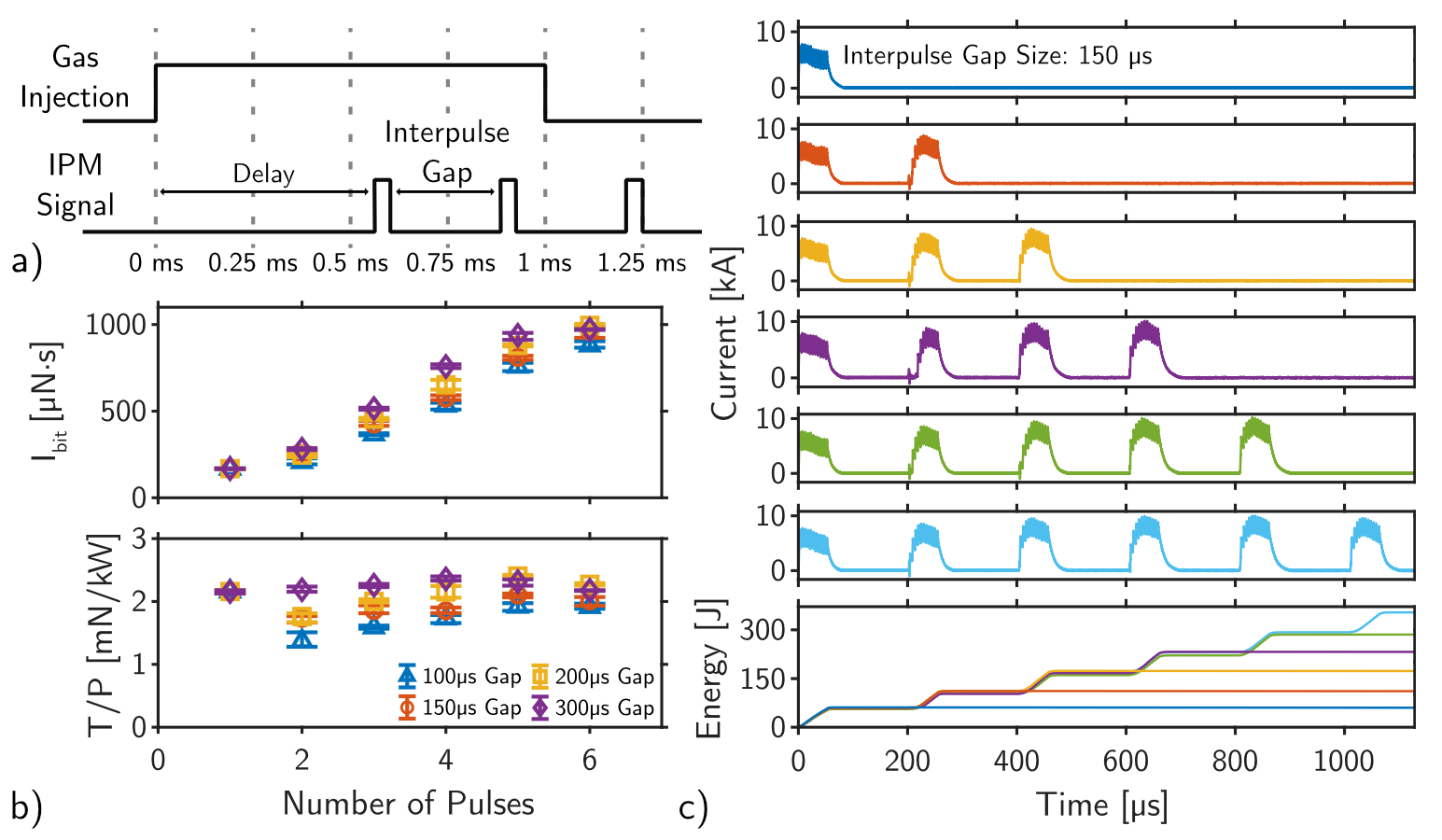}
\caption{Effect of interpulse spacing on micro-burst performance. 
a) Timing diagrams for bursts with varying interpulse gaps. 
b) $I_{\mathrm{bit}}$ increased with pulse count for all gap cases, whereas the evolution of $T/P$ depended strongly on interpulse gap spacing. Shorter gaps produced an initial reduction in $T/P$, while the 300 $\mu$s case remained approximately constant over much of the burst. 
c) Current traces for bursts with a 150 $\mu$s interpulse gap. The second pulse exhibited a slower initial rise and a more rounded profile than the first pulse, while later pulses became progressively more square-like.}
\label{fig:thruster_burst}
\end{figure}

\noindent
To determine how propellant replenishment between discharges influenced impulse accumulation and thrust-to-power ratio, multiple current pulses were applied during a single gas-injection event to form a micro-burst. For each interpulse gap of 100, 150, 200, and 300 $\mu$s, six independent thrust-pendulum measurements were performed using micro-bursts containing one, two, three, four, five, or six pulses (Fig.~\ref{fig:thruster_burst}a). A plenum pressure of 50 psig, a valve-opening duration of 1000 $\mu$s, and a discharge delay of 600 $\mu$s were used for all cases. Each pulse lasted 50 $\mu$s and used an input signal designed to produce an 8 kA current plateau. Current through the accelerator and voltage across the positive and negative flanges were measured to characterize pulse-to-pulse waveform evolution and electrical energy deposition.

The impulse bit increased monotonically with pulse count for all interpulse spacings, whereas the evolution of $T/P$ depended strongly on the time between pulses (Fig.~\ref{fig:thruster_burst}b). For the shorter interpulse gaps, introducing a second pulse produced an immediate decrease in $T/P$, and the magnitude of this reduction increased as the gap was shortened. In contrast, $T/P$ remained approximately constant for the 300 $\mu$s gap, indicating that the accelerator had sufficient time to replenish between successive discharges under these conditions. Because the acceleration and exhaust timescales were much shorter than the propellant-filling timescale, pulses fired with insufficient spacing likely encountered a depleted propellant distribution. These propellant-starved discharges deposited additional energy without producing a proportional increase in impulse. However, excessively long gaps would allow propellant to leave the accelerator before interacting with a subsequent discharge and would reduce the number of pulses that could be applied during a single gas-injection event. Interpulse spacing controls a tradeoff between accelerator replenishment, propellant retention, and the number of useful discharges within each burst.

The approximately constant $T/P$ observed for the 300 $\mu$s case suggests that this spacing provided sufficient time for accelerator replenishment over much of the burst. However, at $N=6$, the 200 $\mu$s case produced an $I_{\mathrm{bit}}$ of approximately 1 mN$\cdot$s, compared with approximately 970 $\mu$N$\cdot$s for the 300 $\mu$s case. By this point in the burst, the valve had been closed for hundreds of microseconds, and the pulses were interacting with a decaying propellant supply. Because the sixth pulse in the 300 $\mu$s sequence occurred later than that in the 200 $\mu$s sequence, this comparison reflects both the time allowed for replenishment and the changing propellant availability after valve closure. These results suggest that a uniform interpulse gap may not maximize performance throughout the full burst. Adaptive pulse timing that accounts for the evolving propellant distribution could improve impulse accumulation and $T/P$, although the present measurements do not establish the optimal progression of interpulse gaps.

\begin{figure}[bp!]
\centering
\includegraphics[width=\figurewidth]{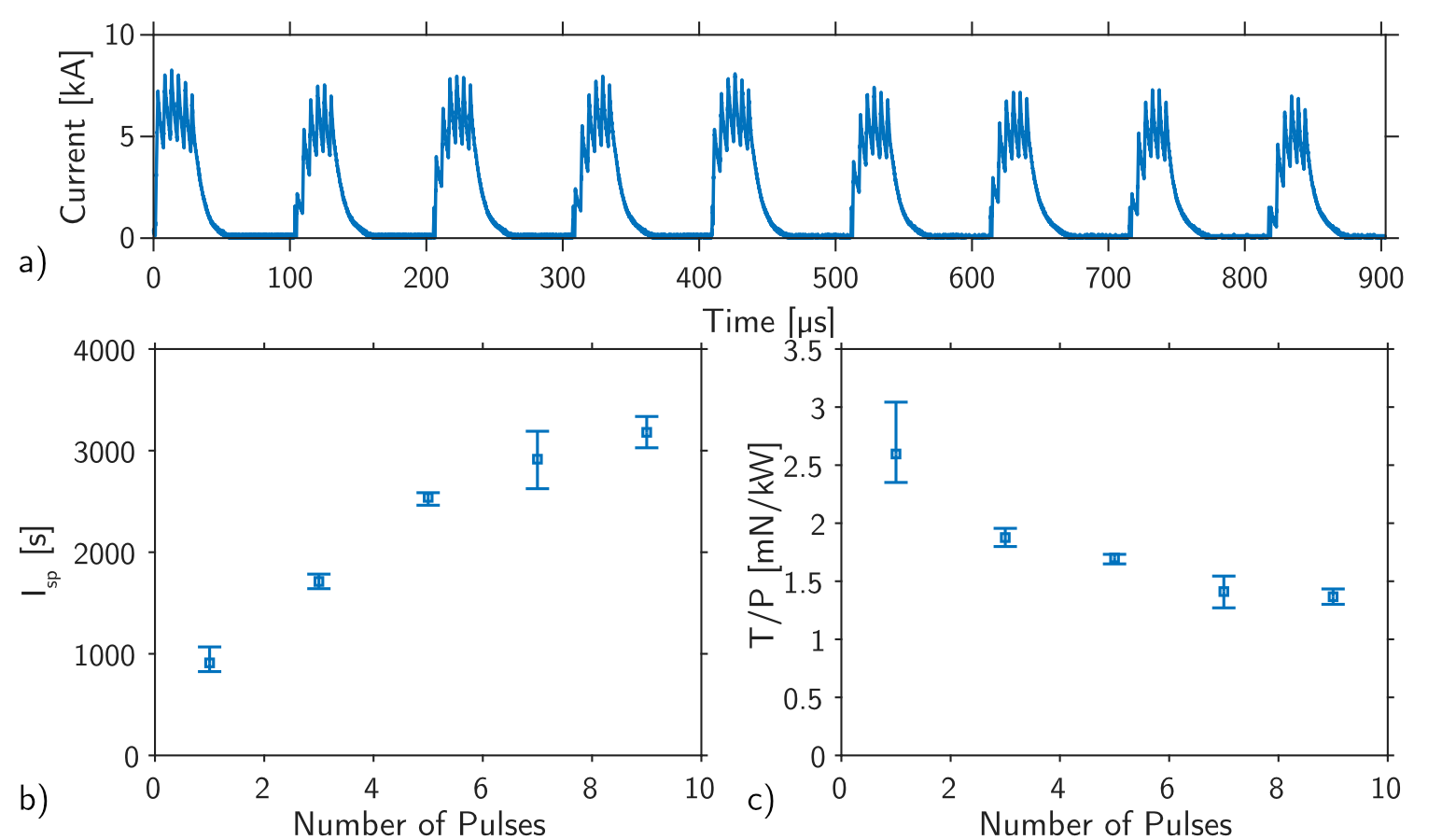}
\caption{Effect of pulse count on the specific impulse obtained from a fixed injected mass bit. 
a) A nine-pulse micro-burst was applied after valve closure using 50 $\mu$s pulses separated by 50 $\mu$s interpulse gaps, extending the total energy addition period to more than 850 $\mu$s. 
b) $I_{\mathrm{sp}}$ increased from 840 to 3177 s as the pulse count increased from one to nine, while the incremental gain diminished at larger pulse counts. 
c) $T/P$ decreased with pulse count, indicating that later pulses produced less impulse per unit energy as the finite propellant inventory in the accelerator was progressively exhausted.}
\label{fig:i_sp}
\end{figure}

Impulse measurements were then performed to determine how much additional impulse could be extracted from a fixed injected mass bit by applying multiple discharges during a single gas-injection event. The thruster was operated at a plenum pressure of 25 psig and a valve-opening duration of 600 $\mu$s, producing an injected mass bit of 12.86 $\mu$g. A delay of 600 $\mu$s was used, and each pulse lasted 50 $\mu$s. The interpulse spacing was set to 50 $\mu$s to maximize the number of discharges within the burst (Fig.~\ref{fig:i_sp}a).

Increasing the pulse count from one to nine raised $I_{\mathrm{sp}}$ from 840 to 3177 s (Fig.~\ref{fig:i_sp}b). Because the injected mass remained constant, this increase resulted directly from the additional impulse accumulated through successive discharges. The incremental increase in $I_{\mathrm{sp}}$ diminished at larger pulse counts. This plateau in $I_{\mathrm{sp}}$ indicated progressive exhaustion of the finite propellant mass remaining in the accelerator. Because all pulses occurred after the valve had closed, no additional propellant was introduced during the burst, and each discharge reduced the mass available to subsequent pulses. At the same time, $T/P$ decreased as pulses were added (Fig.~\ref{fig:i_sp}c), indicating that later discharges deposited energy into an increasingly depleted propellant distribution and produced less impulse per unit energy. Thus, closely packed micro-bursts increased the total impulse extracted from a fixed gas injection, but with diminishing returns and a corresponding reduction in energy-to-impulse conversion.

\subsection{The Expanded Design Space}

\begin{figure}[bp!]
\centering
\includegraphics[width=\figurewidth]{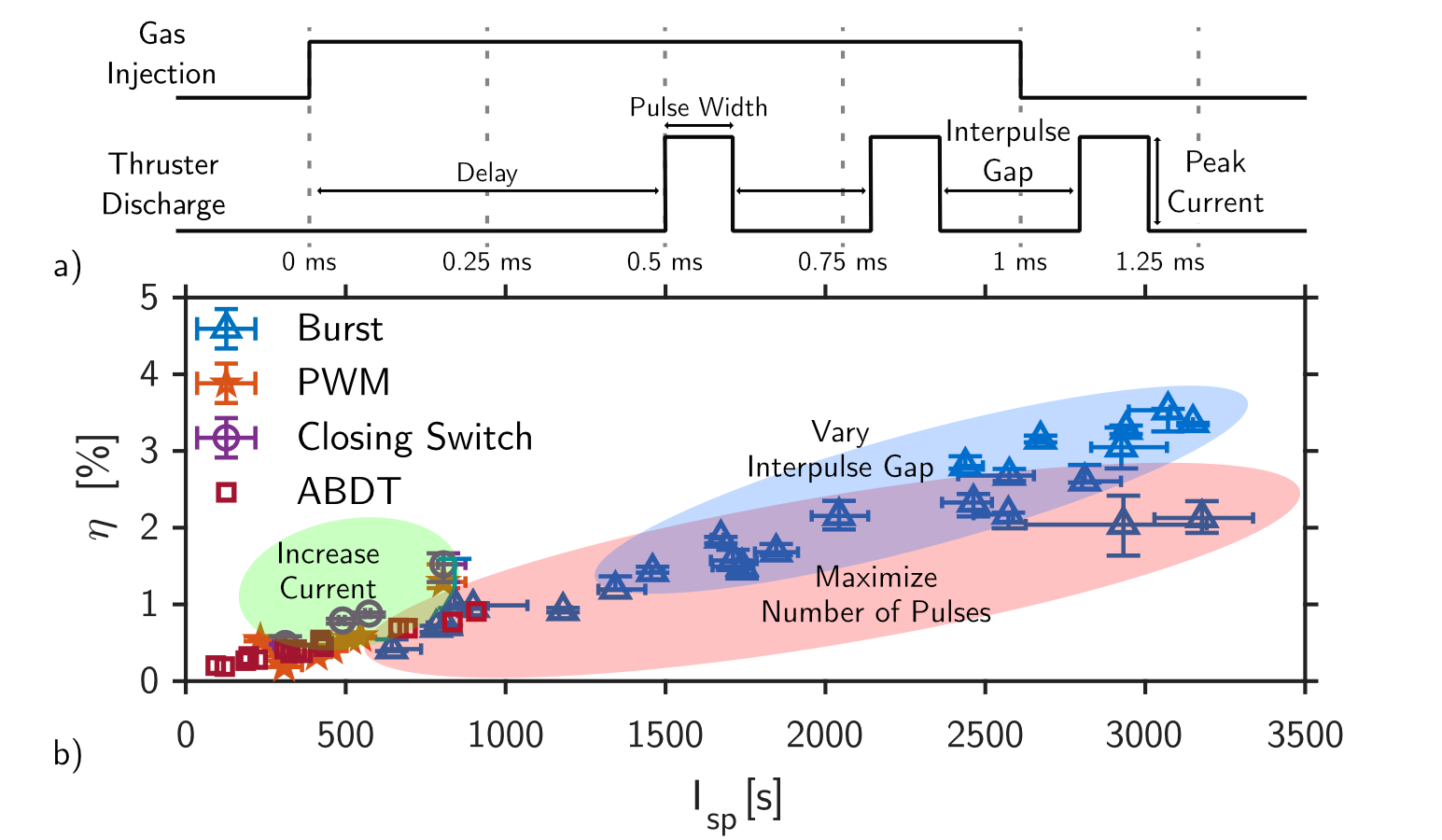}
\caption{Programmable dimensions of the pulse shaping design space and their measured effects on pulsed-electromagnetic-thruster performance.
a) Discharge delay, pulse width, peak current, pulse count, and interpulse spacing can be varied relative to the gas-injection event.
b) The measured operating conditions demonstrate how these controls can be used to navigate the $I_{\mathrm{sp}}$--$\eta$ performance space. Increasing current at fixed pulse width strengthened electromagnetic acceleration, shortening the pulse at fixed energy increased impulse and efficiency, and appropriately timed micro-bursts increased the impulse extracted from a single gas injection. \cite{Subhankar2024}}
\label{fig:comparison}
\end{figure}

\noindent
This paper has demonstrated that programmable pulse shaping expands the accelerator design space by uncoupling peak current from the energy addition timescale. The system provided independent control of discharge delay, pulse duration, peak current, waveform shape, pulse count, interpulse spacing, and pulse-to-pulse energy allocation relative to gas injection. These variables governed the propellant distribution encountered at breakdown, the strength and duration of electromagnetic acceleration, and the distribution of electrical energy across the evolving propellant supply (Fig.~\ref{fig:comparison}a). Their effects can be represented in terms of $I_{\mathrm{sp}}$ and thrust efficiency, $\eta=(g_0/2)I_{\mathrm{sp}}(T/P)$, which respectively characterize impulse per unit injected mass and the conversion of energy into axial momentum.

Varying these control dimensions produced distinct trajectories through the measured $I_{\mathrm{sp}}$--$\eta$ performance space (Fig.~\ref{fig:comparison}b). Discharge delay selected whether acceleration began as a magneto-deflagration or magneto-detonation, while concentrating energy into shorter, higher-current discharges increased plume velocity, radial compression, impulse bit, and thrust efficiency. Micro-bursts extended this approach across the gas-injection event by applying multiple short discharges as the propellant distribution evolved. This produced the largest performance change, increasing $I_{\mathrm{sp}}$ by 278\%, from 840 to 3177 s. Separate interpulse-spacing measurements increased $\eta$ from approximately 0.2\% for a single pulse to 3.3\% for six-pulse bursts with interpulse gaps of 200--300~$\mu$s.

More broadly, programmable pulse shaping allows an electromagnetic thruster to be reconfigured electronically rather than through changes to the accelerator or pulsed-power hardware. By adjusting the timing, duration, amplitude, and distribution of energy addition, a single system could ``shift gears'' among operating conditions that prioritize specific impulse, thrust, efficiency, or available power. This flexibility could reduce the need for propulsion systems specialized for individual applications or mission phases and allow a single accelerator to operate across a broader performance envelope.

\section{Conclusion}
\noindent
This work demonstrated that programmable pulse shaping can transform the current waveform from a fixed consequence of the pulsed-power circuit into an independent design variable. Manipulation of discharge timing, pulse duration, peak current, waveform shape, and burst structure allowed the strength and timing of electromagnetic acceleration to be coordinated with the evolving propellant distribution. This flexibility enabled a single accelerator to access short, high-current GFPPT-like operation, longer pulsed-MPD-like operation, and micro-burst modes without changing the accelerator geometry or pulsed-power hardware. The results establish the electrical waveform and propellant dynamics as coupled design variables that together determine the accessible propulsion performance.

The next generation of pulsed electromagnetic devices can be designed around coordinated control of the electrical waveform and propellant dynamics. Feedback from current, voltage, optical emission, or other plasma diagnostics could ultimately be used to modify pulse timing and waveform shape in real time as the accelerator impedance and mass loading change. This adaptability could allow a single spacecraft to satisfy different propulsion requirements across multiple mission phases, reducing the need to develop and launch separate, highly specialized spacecraft for objectives requiring different propulsion characteristics. Applying this framework across different geometries, propellants, and power levels could produce flexible electromagnetic propulsion systems capable of navigating a broad performance envelope with a single accelerator.

\section*{Appendix}

\appendix

\section{Magnetohydrodynamic Model}
\label{app:mhd_model}

A numerical modeling study of velocity scaling was conducted using a resistive magnetohydrodynamic (MHD) framework with a cell-centered finite-volume formulation in a two-dimensional axisymmetric domain employing a cylindrical coordinate system in the $r$-$z$ plane. The simulations were performed using air. Plasma acceleration was assumed to be axisymmetric, which enabled only the upper half of the thruster to be modeled. The computational domain therefore consisted of an acceleration channel and a plume region and contained approximately 46,000 computational cells. A grid size of $6\times10^{-4}$~m and a time step of $10^{-9}$~s were employed.

The governing equations were spatially discretized using the local Lax-Friedrichs scheme, and the resulting system of ordinary differential equations was advanced in time using a fully implicit backward Euler scheme. The sparse linear systems produced at each implicit time step were solved using the Generalized Minimum Residual (GMRES) Krylov solver implemented through the Portable, Extensible Toolkit for Scientific Computation (PETSc). \cite{Balay2012PETSc3.3} The MHD formulation incorporated finite-rate chemistry and evolved three separate energy equations for the heavy species, electrons, and vibrationally excited molecules. \cite{Baghirzade2026ComputationalSatellites}

The total gas continuity equation was

\begin{equation}
\frac{\partial \rho}{\partial t}
+
\nabla\cdot\left(\rho\mathbf{V}\right)
=
0,
\label{eq:mixture_continuity}
\end{equation}

\noindent
where $\rho$ is the total gas density, $t$ is time, and $\mathbf{V}$ is the bulk flow velocity. Equation~\eqref{eq:mixture_continuity} describes conservation of the total gas mass.

The conservation equation for each independently transported species was

\begin{equation}
\frac{\partial \rho_k}{\partial t}
+
\nabla\cdot\left(\rho_k\mathbf{V}\right)
=
\dot{\omega}_k,
\qquad
\left(k\neq\mathrm{N},e^-\right),
\label{eq:species_continuity}
\end{equation}

\noindent
where $\rho_k$ is the mass density of species $k$ and $\dot{\omega}_k$ is the net mass-production rate of species $k$ per unit volume. The index $k$ identifies the individual chemical species. Equation~\eqref{eq:species_continuity} accounts for convection of each independently transported species with the bulk flow and its production or depletion by finite-rate chemical reactions. In the present formulation, the electron number density was obtained from quasineutrality, while the atomic-nitrogen mass density was recovered from total mass conservation. Species-continuity equations were solved for the remaining seven species.

The momentum equation was

\begin{equation}
\frac{\partial\left(\rho\mathbf{V}\right)}{\partial t}
+
\nabla\cdot
\left(
\rho\mathbf{V}\otimes\mathbf{V}
-
\frac{\mathbf{B}\otimes\mathbf{B}}{\mu_0}
\right)
+
\nabla
\left(
P_h+P_e+\frac{\lvert\mathbf{B}\rvert^2}{2\mu_0}
\right)
=
0,
\label{eq:mhd_momentum}
\end{equation}

\noindent
where $\mathbf{B}$ is the magnetic-field vector, $\mu_0$ is the permeability of free space, $P_h$ is the thermodynamic pressure of the heavy species, and $P_e$ is the electron pressure. Equation~\eqref{eq:mhd_momentum} describes the evolution of the bulk gas momentum under the combined effects of fluid momentum transport, heavy-species and electron pressure, and magnetic stresses.

The magnetic-field evolution equation was

\begin{equation}
\frac{\partial\mathbf{B}}{\partial t}
+
\nabla\cdot
\left(
\mathbf{V}\mathbf{B}
-
\mathbf{B}\mathbf{V}
\right)
=
-\nabla\times
\left(
\eta_e\frac{\nabla\times\mathbf{B}}{\mu_0}
\right),
\label{eq:magnetic_field}
\end{equation}

\noindent
where $\eta_e$ is the plasma electrical resistivity. The products $\mathbf{V}\mathbf{B}$ and $\mathbf{B}\mathbf{V}$ are dyadic products. The left-hand side of Eq.~\eqref{eq:magnetic_field} describes temporal evolution and transport of the magnetic field by the bulk flow, while the right-hand side accounts for resistive diffusion of the magnetic field.

The heavy-species total-energy equation was

\begin{equation}
\frac{\partial\left(\rho e_t^h\right)}{\partial t}
+
\nabla\cdot
\left[
\left(
\rho e_t^h+P_h+P_e
\right)\mathbf{V}
\right]
=
\mathbf{V}\cdot\left(\mathbf{J}\times\mathbf{B}\right)
+
\Omega^{ET}
+
P_e\nabla\cdot\mathbf{V}
+
S_{\mathrm{chem}}^h,
\label{eq:heavy_energy}
\end{equation}

\noindent
where $e_t^h$ is the heavy-species total specific energy, $\mathbf{J}$ is the current-density vector, and $\times$ denotes the vector cross product. The term $\mathbf{J}\times\mathbf{B}$ is the Lorentz-force density, and $\mathbf{V}\cdot(\mathbf{J}\times\mathbf{B})$ is the corresponding mechanical work performed on the gas. The term $\Omega^{ET}$ represents translational energy exchange between the electrons and heavy species, $P_e\nabla\cdot\mathbf{V}$ represents electron-pressure work transferred to the heavy-species energy mode, and $S_{\mathrm{chem}}^h$ represents chemical energy exchange with the heavy-species energy mode. Equation~\eqref{eq:heavy_energy} describes the evolution and transport of the heavy-species total energy.

The electron-energy equation was

\begin{equation}
\frac{\partial\left(\rho_e e_e\right)}{\partial t}
+
\nabla\cdot
\left[
\left(\rho_e e_e\right)\mathbf{V}
\right]
=
\eta_e J^2
-
\Omega^{ET}
+
S_{\mathrm{chem}}^e
-
P_e\nabla\cdot\mathbf{V},
\label{eq:electron_energy}
\end{equation}

\noindent
where $\rho_e$ is the electron mass density, and $e_e$ is the electron specific energy. The term $\eta_eJ^2$ represents resistive Joule heating of the electrons. The term $-\Omega^{ET}$ represents energy transferred from the electrons to the heavy species, $S_{\mathrm{chem}}^e$ represents chemical energy exchange with the electron-energy mode, and $-P_e\nabla\cdot\mathbf{V}$ represents electron compression or expansion work. Equation~\eqref{eq:electron_energy} describes the evolution and convection of electron energy.

The vibrational-energy equation was

\begin{equation}
\frac{\partial\left(\rho_v e_v\right)}{\partial t}
+
\nabla\cdot
\left[
\left(\rho_v e_v\right)\mathbf{V}
\right]
=
\Omega^{EV}
+
\Omega^{VT}
+
S_{\mathrm{chem}}^{\mathrm{vib}},
\label{eq:vibrational_energy}
\end{equation}

\noindent
where $\rho_v e_v$ is the vibrational energy per unit volume. The term $\Omega^{EV}$ represents the change in vibrational energy associated with electron-impact vibrational excitation reactions, and $\Omega^{VT}$ represents collisional relaxation between the vibrational energy mode and the thermal energy of the heavy species. The term $S_{\mathrm{chem}}^{\mathrm{vib}}$ accounts for changes in vibrational energy caused by the finite-rate formation and depletion of molecular species. Equation~\eqref{eq:vibrational_energy} describes the evolution and convection of the molecular vibrational energy.

The finite-rate air-chemistry model included nine species,

\begin{equation}
\mathrm{
N_2,\,
O_2,\,
N,\,
O,\,
N_2^+,\,
O_2^+,\,
N^+,\,
O^+,\,
e^-
},
\label{eq:air_species}
\end{equation}

\noindent
where $\mathrm{N_2}$ and $\mathrm{O_2}$ are molecular nitrogen and oxygen, $\mathrm{N}$ and $\mathrm{O}$ are atomic nitrogen and oxygen, $\mathrm{N_2^+}$ and $\mathrm{O_2^+}$ are the corresponding molecular ions, $\mathrm{N^+}$ and $\mathrm{O^+}$ are the corresponding atomic ions, and $e^-$ denotes electrons.

The plasma electrical resistivity was evaluated according to

\begin{equation}
\eta_e
=
\frac{m_e\left(\nu_{en}+\nu_{ei}\right)}
{n_e e^2},
\label{eq:resistivity}
\end{equation}

\noindent
where $m_e$ is the electron mass, $\nu_{en}$ is the electron-neutral collision frequency, $\nu_{ei}$ is the electron-ion collision frequency, $n_e$ is the electron number density, and $e$ is the elementary charge. Equation~\eqref{eq:resistivity} relates the plasma resistivity to electron momentum loss through collisions with neutral particles and ions.

The electron-neutral collision frequency was calculated as

\begin{equation}
\nu_{en}
=
\sum_k
n_k\sigma_k
\sqrt{\frac{8k_BT_e}{\pi m_e}},
\label{eq:electron_neutral_collision}
\end{equation}

\noindent
where $n_k$ is the number density of background neutral species $k$, $\sigma_k$ is the corresponding electron-neutral collision cross section, $k_B$ is the Boltzmann constant, $T_e$ is the electron temperature, and $\pi$ is the circle constant. The summation is taken over the background neutral species included in the collision model. The square-root term represents the electron thermal-speed dependence of the collision frequency. The collision cross sections were obtained from the LXCat database.

The electron-ion collision frequency was calculated as

\begin{equation}
\nu_{ei}
=
3.636\times10^{-6}
n_eT_e^{-3/2}
\ln\left(\Lambda\right),
\label{eq:electron_ion_collision}
\end{equation}

\noindent
where $\ln(\Lambda)$ is the Coulomb logarithm and the remaining quantities are defined above. Equation~\eqref{eq:electron_ion_collision} relates the electron-ion collision frequency to the electron number density and electron temperature. Additional details of the numerical framework are provided in Ref.~\cite{Baghirzade2026ComputationalSatellites}.

To replicate the vacuum-arc discharge observed in the experiments, a thermal-plasma inflow boundary condition was imposed in the simulations. The inlet temperature and pressure were prescribed as

\begin{equation}
T_{\mathrm{in}}=10{,}500~\mathrm{K},
\qquad
P_{\mathrm{in}}=110~\mathrm{Pa},
\label{eq:inflow_condition}
\end{equation}

\noindent
where $T_{\mathrm{in}}$ is the prescribed inlet temperature and $P_{\mathrm{in}}$ is the prescribed inlet pressure. These values yield agreement with the nominal exhaust velocity, stagnation pressure, and plasma density from experiments. \cite{Subramaniam2018ComputationalAccelerators,Underwood2019b} The inlet density was obtained from Mutation++ using these prescribed temperature and pressure values. \cite{Scoggins2020Mutation++:C++}

The applied current was modeled at the inlet by imposing the corresponding azimuthal magnetic field. For the coaxial accelerator, the azimuthal magnetic field was obtained by integrating Amp\`ere's law over the circular cross section,

\begin{equation}
B_\theta(r,t)
=
\frac{\mu_0I(t)}{2\pi r},
\label{eq:imposed_magnetic_field}
\end{equation}

\noindent
where $B_\theta$ is the azimuthal component of the magnetic field, $I(t)$ is the prescribed time-dependent discharge current, and $r$ is the radial distance from the accelerator centerline. The peak inlet magnetic-field strengths corresponding to applied currents of 10, 15, 20, and 25~kA were approximately 0.08, 0.12, 0.153, and 0.20~T, respectively. Increasing the applied current strengthened the Lorentz acceleration and consequently increased the exhaust velocity, $V_{\mathrm{ex}}$.

The temporal evolution of the effective exhaust velocity was evaluated at the accelerator exit plane using a spatial, mass-flux-weighted average,

\begin{equation}
V_{\mathrm{ex}}(t)
=
\frac{
\displaystyle
\iint_A
\rho V_z
\left(
\mathbf{V}\cdot\hat{\mathbf{n}}
\right)
\,dA
}{
\displaystyle
\iint_A
\rho
\left(
\mathbf{V}\cdot\hat{\mathbf{n}}
\right)
\,dA
},
\label{eq:mass_weighted_velocity_general}
\end{equation}

\noindent
where $V_{\mathrm{ex}}(t)$ is the time-dependent effective exhaust velocity, $A$ is the accelerator exit-plane area, $V_z$ is the axial velocity, $\hat{\mathbf{n}}$ is the unit vector normal to the exit plane, and $\mathbf{V}\cdot\hat{\mathbf{n}}$ is the velocity component normal to the exit plane. The differential area is $dA=r\,dr\,d\theta$, where $\theta$ is the azimuthal coordinate. The numerator is the axial-momentum flux through the exit plane, while the denominator is the mass flow rate through the same plane.

Because the exit-plane normal was aligned with the axial direction, $\hat{\mathbf{n}}=\hat{\mathbf{z}}$ and $\mathbf{V}\cdot\hat{\mathbf{n}}=V_z$. Under the axisymmetric formulation, Eq.~\eqref{eq:mass_weighted_velocity_general} becomes

\begin{equation}
V_{\mathrm{ex}}(t)
=
\frac{
\displaystyle
\int
\rho V_z^2\,2\pi r\,dr
}{
\displaystyle
\int
\rho V_z\,2\pi r\,dr
}.
\label{eq:mass_weighted_velocity}
\end{equation}

\noindent
Here, $2\pi r\,dr$ is the differential annular area obtained by integrating over the azimuthal coordinate. Equation~\eqref{eq:mass_weighted_velocity} gives the mass-flux-weighted axial velocity across the accelerator exit plane.

The total electromagnetic power deposited into the plasma was calculated by integrating $\mathbf{J}\cdot\mathbf{E}$ over the acceleration-channel control volume,

\begin{equation}
P_{\mathrm{EM}\rightarrow\mathrm{plasma}}(t)
=
\iiint
\left(\mathbf{J}\cdot\mathbf{E}\right)dV
=
\iint
\left[
\eta_e J^2
+
\mathbf{V}\cdot(\mathbf{J}\times\mathbf{B})
\right]
2\pi r\,dS,
\end{equation}

\noindent
where $dV=r\,dr\,dz\,d\theta$ and $dS=dr\,dz$. The deposited electromagnetic power is partitioned into resistive Joule heating, $\eta_e J^2$, and Lorentz mechanical work, $\mathbf{V}\cdot(\mathbf{J}\times\mathbf{B})$. The ratio of Lorentz mechanical work to Joule heating reported in Section~2.1 was calculated from the quasi-steady values of these volume-integrated power terms.

\section*{Declarations}

\begin{flushleft}
\setlength{\parindent}{0pt}

\textbf{Funding}  
This research is supported by the Defense Advanced Research Projects Agency TALOS program award UTAUS-FA00003150 and the NASA ECF Program award 80NSSC25K7834.

\textbf{Conflicts of interest/Competing interests}  
The authors declare no competing interests.

\textbf{Data availability}  
The datasets generated and analyzed during the current study are available from the corresponding author on reasonable request.

\textbf{Code availability}  
Code used for analysis is available from the corresponding author on reasonable request.

\textbf{Authors' contributions}  
P. Schools constructed the pulsed power supply, conducted experiments, analyzed data, and wrote the manuscript. M. Baghirzade performed the numerical modeling and contributed to the manuscript. R. Heiser processed thrust data. A. Woodley aided in thrust data collection. L. Raja provided mentorship. T. Underwood conceived the study and contributed to experimental design, analysis, and manuscript review.

\end{flushleft}

\bibliography{References/PulseShapingJEP}

@article{Keidar2004,
   author = {M Keidar and I D Boyd and I I Beilis},
   doi = {10.1063/1.1805726},
   issn = {0021-8979},
   issue = {10},
   journal = {Journal of Applied Physics},
   month = {11},
   pages = {5420-5428},
   title = {Ionization and ablation phenomena in an ablative plasma accelerator},
   volume = {96},
   year = {2004}
}

@article{Loebner2015,
   author = {Keith T K Loebner and Thomas C Underwood and Mark A Cappelli},
   doi = {10.1063/1.4922522},
   issn = {0034-6748},
   issue = {6},
   journal = {Review of Scientific Instruments},
   month = {6},
   pages = {063503},
   title = {A fast rise-rate, adjustable-mass-bit gas puff valve for energetic pulsed plasma experiments},
   volume = {86},
   year = {2015}
}

@article{Underwood2019a,
   author = {Thomas C Underwood and Keith T K Loebner and Victor A Miller and Mark A Cappelli},
   doi = {10.1007/s00348-019-2848-5},
   issn = {1432-1114},
   issue = {1},
   journal = {Experiments in Fluids},
   pages = {17},
   title = {Schlieren diagnostic for cinematic visualization of dense plasma jets at Alfvénic timescales},
   volume = {61},
   year = {2019}
}

@article{Woodley2025b,
   author = {Adrian Woodley and Ethan Horstman and Thomas C Underwood},
   doi = {10.1063/5.0273140},
   issn = {0021-8979},
   issue = {6},
   journal = {Journal of Applied Physics},
   month = {8},
   pages = {063303},
   title = {Magnetohydrodynamic operating regimes of pulsed plasma accelerators for efficient propellant utilization},
   volume = {138},
   year = {2025}
}

@article{Underwood2021,
   author = {Thomas C Underwood and William M Riedel and Mark A Cappelli},
   doi = {10.1063/5.0051467},
   issn = {0021-8979},
   issue = {13},
   journal = {Journal of Applied Physics},
   month = {10},
   pages = {133301},
   title = {Dual mode operation of a hydromagnetic plasma thruster to achieve tunable thrust and specific impulse},
   volume = {130},
   year = {2021}
}

@article{Subhankar2024,
   author = {Varanasi Sai Subhankar and Keshav P Prathivadi and Thomas C Underwood},
   doi = {https://doi.org/10.1016/j.actaastro.2023.12.048},
   issn = {0094-5765},
   journal = {Acta Astronautica},
   pages = {91-101},
   title = {Deflagration thruster for air-breathing electric propulsion in very low Earth orbit},
   volume = {216},
   url = {https://www.sciencedirect.com/science/article/pii/S0094576523006781},
   year = {2024}
}

@article{Underwood2019b,
   author = {Thomas C Underwood and Vivek Subramaniam and William M Riedel and Laxminarayan L Raja and Mark A Cappelli},
   doi = {https://doi.org/10.1016/j.fusengdes.2019.04.088},
   issn = {0920-3796},
   journal = {Fusion Engineering and Design},
   pages = {97-106},
   title = {{Effects of flow collisionality on ELM replication in plasma guns}},
   volume = {144},
   url = {https://www.sciencedirect.com/science/article/pii/S0920379619306283},
   year = {2019}
}

@article{Underwood2019c,
   author = {Thomas C Underwood and Keith T K Loebner and Victor A Miller and Mark A Cappelli},
   doi = {10.1038/s41598-019-39827-6},
   issn = {2045-2322},
   issue = {1},
   journal = {Scientific Reports},
   pages = {2588},
   title = {Dynamic formation of stable current-driven plasma jets},
   volume = {9},
   year = {2019}
}

@article{Underwood2017,
   author = {Thomas C Underwood and Keith T K Loebner and Mark A Cappelli},
   doi = {https://doi.org/10.1016/j.hedp.2017.03.004},
   issn = {1574-1818},
   journal = {High Energy Density Physics},
   pages = {73-80},
   title = {A plasma deflagration accelerator as a platform for laboratory astrophysics},
   volume = {23},
   url = {https://www.sciencedirect.com/science/article/pii/S1574181817300204},
   year = {2017}
}

@article{Loebner2016,
   author = {Keith T K Loebner and Thomas C Underwood and Theodore Mouratidis and Mark. A Cappelli},
   doi = {10.1063/1.4943370},
   issn = {0003-6951},
   issue = {9},
   journal = {Applied Physics Letters},
   month = {3},
   pages = {094104},
   title = {Radial magnetic compression in the expelled jet of a plasma deflagration accelerator},
   volume = {108},
   year = {2016}
}

@article{Loebner2015Branch,
   author = {Keith T.K. Loebner and Thomas C Underwood and Mark A Cappelli},
   doi = {10.1103/PhysRevLett.115.175001},
   issue = {17},
   journal = {Physical Review Letters},
   month = {10},
   pages = {175001},
   publisher = {American Physical Society},
   title = {Evidence of Branching Phenomena in Current-Driven Ionization Waves},
   volume = {115},
   year = {2015}
}

@inproceedings{Johnson2002,
  author    = {Johnson, Les and Jones, Jonathan and Kos, Larry and
               Trausch, Ann and Farris, Robert and Woodcock, Gordon},
  title     = {Benefits of Nuclear Electric Propulsion to Outer Planets Exploration},
  booktitle = {38th AIAA/ASME/SAE/ASEE Joint Propulsion Conference \& Exhibit},
  publisher = {American Institute of Aeronautics and Astronautics},
  address   = {Reston, VA},
  month     = jul,
  year      = {2002},
  doi       = {10.2514/6.2002-3548}
}

@techreport{Choueiri2000,
    author      = {Choueiri, Edgar Y.},
    title       = {Gas-Fed Pulsed Plasma Thrusters: Fundamentals, Characteristics and Scaling Laws},
    year        = {2000},
    month       = {12},
    institution = {Princeton University},
    type        = {Final Report},
    number      = {F49620-98-1-0119},
    address     = {Princeton, New Jersey}
}

@article{LaPointe2004,
   author = {Michael LaPointe and Eugene Strzempkowski and Eric Pencil},
   doi = {10.2514/6.2004-3467},
   journal = {40th AIAA/ASME/SAE/ASEE Joint Propulsion Conference and Exhibit},
   month = {10},
   title = {High Power MPD Thruster Performance Measurements},
   year = {2004}
}

@article{Huang2020,
   author = {Tiankun Huang and Zhiwen Wu and Guorui Sun and Xiangyang Liu and William Yeong Liang Ling},
   doi = {https://doi.org/10.1016/j.actaastro.2020.04.010},
   issn = {0094-5765},
   journal = {Acta Astronautica},
   pages = {69-75},
   title = {Study and modeling of propellant ablation in coaxial ablative pulsed plasma thrusters},
   volume = {173},
   url = {https://www.sciencedirect.com/science/article/pii/S0094576520302149},
   year = {2020}
}

@article{Ziemer2000,
   author = {John Ziemer and Edgar Choueiri},
   doi = {10.2514/6.2000-3432},
   journal = {36th AIAA/ASME/SAE/ASEE Joint Propulsion Conference and Exhibit},
   month = {7},
   title = {A Characteristic Velocity for Gas-Fed PPT Performance Scaling},
   year = {2000}
}

@article{Organski2025,
   author = {Lee Organski and Brian Jeffers and Patrick Gresham and Artur Kucharewicz and Alexey Shashurin},
   doi = {10.1063/5.0242934},
   issn = {2158-3226},
   issue = {2},
   journal = {AIP Advances},
   month = {2},
   pages = {025007},
   title = {Low-voltage operation mode of ASCENT-propelled pulsed plasma thruster},
   volume = {15},
   year = {2025}
}

@article{Choueiri2001,
   author = {Edgar Choueiri and John Ziemer},
   doi = {10.2514/2.5857},
   journal = {Journal of Propulsion and Power},
   month = {9},
   pages = {967-976},
   title = {Quasi-Steady Magnetoplasmadynamic Thruster Performance Database},
   volume = {17},
   year = {2001}
}

@article{Domonkos1995,
   author = {Matt Domonkos and Alec Gallimore and Roger Myers},
   doi = {10.2514/6.1995-2674},
   month = {7},
   title = {Preliminary Pulsed MPD Thruster Performance},
   year = {1995}
}

@techreport{Balay2012PETSc3.3,
    title = {{PETSc Users Manual, Revision 3.3}},
    year = {2012},
    author = {Balay, Satish and Brown, Jed and Buschelman, Kris and Eijkhout, Victor and Gropp, William D. and Kaushik, Dinesh and Knepley, Matthew G. and McInnes, Lois Curfman and Smith, Barry F. and Zhang, Hong},
    institution = {Computer Science Division, Argonne National Laboratory},
    address = {Argonne, IL}
}

@phdthesis{Baghirzade2026ComputationalSatellites,
    title = {{Computational modeling of plasma discharges in air-breathing electric propulsion devices for very low Earth orbit satellites}},
    year = {2026},
    author = {Baghirzade, Mammadbaghir},
    school = {The University of Texas at Austin},
    address = {Austin}
}

@article{Scoggins2020Mutation++:C++,
    title = {{Mutation++: MUlticomponent Thermodynamic And Transport properties for IONized gases in C++}},
    year = {2020},
    journal = {SoftwareX},
    author = {Scoggins, James B and Leroy, Vincent and Bellas-Chatzigeorgis, Georgios and Dias, Bruno and Magin, Thierry E},
    pages = {100575},
    volume = {12},
    url = {https://www.sciencedirect.com/science/article/pii/S2352711020302880},
    doi = {https://doi.org/10.1016/j.softx.2020.100575},
    issn = {2352-7110}
}

@article{Marchioni2021ExtendedPropulsion,
    title = {{Extended channel Hall thruster for air-breathing electric propulsion}},
    year = {2021},
    journal = {Journal of Applied Physics},
    author = {Marchioni, Francesco and Cappelli, Mark A.},
    number = {5},
    month = {8},
    volume = {130},
    publisher = {American Institute of Physics Inc.},
    doi = {10.1063/5.0048283},
    issn = {10897550}
}

@inproceedings{BURTON1981,
  author    = {Burton, R. L. and Clark, K. E. and Jahn, R. G.},
  title     = {Thrust and Efficiency of a Self-Field {MPD} Thruster},
  booktitle = {15th International Electric Propulsion Conference},
  address   = {Las Vegas, Nevada},
  month     = apr,
  year      = {1981},
  number    = {AIAA 81-0684},
  publisher = {American Institute of Aeronautics and Astronautics},
  doi       = {10.2514/6.1981-684}
}

@techReport{Castillo1991,
   author = {Salvador Castillo},
   city = {Edwards Airforce Base},
   institution = {Phillips Laboratory},
   month = {12},
   title = {{Establishment of MPD Performance}},
   year = {1991}
}

@article{York1993DiagnosticsNozzle,
    title = {{Diagnostics and Performance of a Low-Power MPD Thruster with Applied Magnetic Nozzle}},
    year = {1993},
    journal = {Journal of Propulsion and Power},
    author = {York, T. M. and Zakrzwski, C. and Soulas, G.},
    number = {4},
    pages = {553--560},
    volume = {9},
    doi = {10.2514/3.23658},
    issn = {07484658}
}

@article{Toki2000On-orbitSystem,
    title = {{On-orbit demonstration of a pulsed self-field magnetoplasmadynamic thruster system}},
    year = {2000},
    journal = {Journal of Propulsion and Power},
    author = {Toki, K. and Shimizu, Y. and Kuriki, K.},
    number = {5},
    pages = {880--886},
    volume = {16},
    publisher = {AIAA},
    doi = {10.2514/2.5655},
    issn = {07484658}
}

@inproceedings{Ziemer1999IsPerformance,
  author    = {Ziemer, Joseph K. and Choueiri, Edgar Y. and Birx, Daniel},
  title     = {Is the Gas-Fed {PPT} an Electromagnetic Accelerator?
               An Investigation Using Measured Performance},
  booktitle = {35th Joint Propulsion Conference and Exhibit},
  address   = {Los Angeles, California},
  month     = jun,
  year      = {1999},
  number    = {AIAA 99-2289},
  publisher = {American Institute of Aeronautics and Astronautics},
  doi       = {10.2514/6.1999-2289}
}

@phdthesis{Ziemer2001,
   author = {John Kenneth Ziemer},
   school = {Princeton University},
   month = {6},
   title = {Performance Scaling of Gas-Fed Pulsed Plasma Thrusters},
   year = {2001}
}

@article{Cheng1971ApplicationThruster,
    title = {{Application of a deflagration plasma gun as a space propulsion thruster}},
    year = {1971},
    journal = {AIAA Journal},
    author = {Cheng, Dah Yu},
    number = {9},
    pages = {1681--1685},
    volume = {9},
    doi = {10.2514/3.6418},
    issn = {00011452}
}

@misc{Zimmerman2025PulsedOrbit,
    title = {{Pulsed Magnetoplasmadynamic Propulsion for Airbreathing Satellites in Very Low Earth Orbit}},
    year = {2025},
    booktitle = {Journal of Propulsion and Power},
    author = {Zimmerman, Joseph W. and Burton, Rodney L. and Fox, Ryan T. and Carroll, David L. and Broemmelsiek, Emil J. and Choueiri, Edgar},
    number = {6},
    month = {11},
    pages = {810--813},
    volume = {41},
    publisher = {AIAA International},
    doi = {10.2514/1.B39990},
    issn = {15333876}
}

@techreport{ZiemerAHalf-Century,
    author      = {Ziemer, John K.},
    title       = {A Review of Gas-Fed Pulsed Plasma Thruster Research over the Last Half-Century},
    year        = {2000},
    institution = {Electric Propulsion and Plasma Dynamics Laboratory, Princeton University},
    address     = {Princeton, New Jersey},
    type        = {Technical Report}
}

@inproceedings{Hauze1992EffectPerformance,
    title = {{Effect of anode size on deflagration accelerator performance}},
    year = {1992},
    address   = {Nashville, Tennessee},
    booktitle = {AIAA 23rd Plasmadynamics and Lasers Conference, 1992},
    author = {Hauze, Gene E. and Green, James E. and Wallace, Richard J.},
    publisher = {American Institute of Aeronautics and Astronautics Inc, AIAA},
    doi = {10.2514/6.1992-3033}
}

@inproceedings{Ziemer1997PerformanceThruster,
    title = {{Performance characterization of a high efficiency gas-fed pulsed plasma thruster}},
    year = {1997},
    address   = {Seattle, Washington},
    booktitle = {33rd Joint Propulsion Conference and Exhibit},
    author = {Ziemer, J. K. and Cubbin, E. A. and Choueiri, E. Y. and Birx, Daniel},
    publisher = {American Institute of Aeronautics and Astronautics Inc, AIAA},
    doi = {10.2514/6.1997-2925}
}

@article{Winands2005LongApplications,
    title = {{Long lifetime, triggered, spark-gap switch for repetitive pulsed power applications}},
    year = {2005},
    journal = {Review of Scientific Instruments},
    author = {Winands, G. J.J. and Liu, Z. and Pemen, A. J.M. and Van Heesch, E. J.M. and Yan, K.},
    number = {8},
    month = {8},
    pages = {1--6},
    volume = {76},
    doi = {10.1063/1.2008047},
    issn = {00346748}
}

@article{Laya2026AGun,
    title = {{A Thyristor-Switched Pulsed Power System With 78-ns Jitter for a Coaxial Plasma Gun}},
    year = {2026},
    journal = {IEEE Transactions on Plasma Science},
    author = {Laya, Neil P. and Tymoshevska, Tetiana and Chesny, David L. and Moffett, Mark B. and Boehm, Kirk J. and Xu, Kunning G.},
    month = {6},
    publisher = {Institute of Electrical and Electronics Engineers Inc.},
    doi = {10.1109/TPS.2026.3692821},
    issn = {19399375}
}

@article{Promislow2022OperationRates,
    title = {{Operation and Performance of a Power Processing Unit for Inductive Pulsed Plasma Thrusters Operating at High Repetition Rates}},
    year = {2022},
    journal = {IEEE Transactions on Plasma Science},
    author = {Promislow, Curtis and Little, Justin},
    number = {9},
    month = {9},
    pages = {3065--3076},
    volume = {50},
    publisher = {Institute of Electrical and Electronics Engineers Inc.},
    doi = {10.1109/TPS.2022.3189678},
    issn = {19399375}
}

@misc{Polzin2020State-of-the-artThrusters,
    title = {{State-of-the-art and advancement paths for inductive pulsed plasma thrusters}},
    year = {2020},
    booktitle = {Aerospace},
    author = {Polzin, Kurt and Martin, Adam and Little, Justin and Promislow, Curtis and Jorns, Benjamin and Woods, Joshua},
    number = {8},
    month = {8},
    volume = {7},
    publisher = {MDPI Multidisciplinary Digital Publishing Institute},
    doi = {10.3390/AEROSPACE7080105},
    issn = {22264310}
}

@misc{Zhang2019AThrusters,
    title = {{A review of the characterization and optimization of ablative pulsed plasma thrusters}},
    year = {2019},
    booktitle = {Reviews of Modern Plasma Physics},
    author = {Zhang, Zhe and Ling, William Yeong Liang and Tang, Haibin and Cao, Jinbin and Liu, Xiangyang and Wang, Ningfei},
    number = {1},
    month = {12},
    volume = {3},
    publisher = {Springer},
    doi = {10.1007/s41614-019-0027-z},
    issn = {23673192}
}

@inproceedings{Molina-CabreraPPulsedClassification,
    title = {{Pulsed Plasma Thrusters: a worldwide review and long yearned classification}},
    year = {2011},
    month = {9},
    author = {{Molina-Cabrera P} and {Herdrich G} and {Lau M} and {Fausolas S} and {Schoenherr T} and {Komurasaki K}},
    booktitle={32nd International Electric Propulsion Conference, IEPC-2011-340},
}

@phdthesis{Crandall2024,
    title = {{Miniature Rf Gridded Ion Thruster for Air-Breathing and Alternative Propellants}},
    year = {2024},
    author = {Crandall, Patrick},
    school = {University of California Los Angeles}
}

@article{Woodall1985ObservationDeflagration,
    title = {{Observation of current sheath transition from snowplow to deflagration}},
    year = {1985},
    journal = {Journal of Applied Physics},
    author = {Woodall, D. M. and Len, L. K.},
    number = {3},
    pages = {961--964},
    volume = {57},
    doi = {10.1063/1.334697},
    issn = {00218979}
}

@techreport{Ducati1971InvestigationJets,
    title = {{Investigation of pulsed quasi-steady MPD arc jets}},
    year = {1971},
    author = {Ducati, A. C. and Jahn, R. G.},
    month = {6},
    institution = {National Aeronautics and Space Administration}
}

@article{Poehlmann2010CurrentMode,
    title = {{Current distribution measurements inside an electromagnetic plasma gun operated in a gas-puff mode}},
    year = {2010},
    journal = {Physics of Plasmas},
    author = {Poehlmann, Flavio R and Cappelli, Mark A and Rieker, Gregory B},
    number = {12},
    month = {12},
    pages = {123508},
    volume = {17},
    url = {https://doi.org/10.1063/1.3526603},
    doi = {10.1063/1.3526603},
    issn = {1070-664X}
}

@article{Huang2015StudyThruster,
    title = {{Study of breakdown in an ablative pulsed plasma thruster}},
    year = {2015},
    journal = {Physics of Plasmas},
    author = {Huang, Tiankun and Wu, Zhiwen and Liu, Xiangyang and Xie, Kan and Wang, Ningfei and Cheng, Yue},
    number = {10},
    month = {10},
    pages = {103511},
    volume = {22},
    url = {https://doi.org/10.1063/1.4933211},
    doi = {10.1063/1.4933211},
    issn = {1070-664X}
}

@inproceedings{Ziemba2011EHTIGBT,
  author    = {Ziemba, Timothy and Miller, Kenneth E. and Prager, James and Carscadden, John},
  title     = {A Robust, Modular, {IGBT} Power Supply for Configurable Series/Parallel Operation at High Power and Frequency},
  booktitle = {2011 IEEE Pulsed Power Conference},
  year      = {2011},
  address   = {Chicago, IL, USA},
  doi       = {10.1109/PPC.2011.6191607}
}

@misc{Miller2013EHTIPM,
  author       = {Miller, Kenneth E.},
  title        = {The {EHT} Integrated Power Module ({IPM}): An {IGBT}-Based, High Current, Ultra-Fast, Modular, Programmable Power Supply Unit},
  year         = {2013},
  howpublished = {Eagle Harbor Technologies technical presentation}
}

@article{Woodley2024RequirementsOrbits,
    title = {{Requirements for air-breathing electric propulsion in low-altitude orbits}},
    year = {2024},
    journal = {Journal of Electric Propulsion},
    author = {Woodley, Adrian and Horstman, Ethan and Keidar, Michael and Underwood, Thomas C},
    number = {1},
    pages = {34},
    volume = {3},
    url = {https://doi.org/10.1007/s44205-024-00095-w},
    doi = {10.1007/s44205-024-00095-w},
    issn = {2731-4596}
}

@article{Benkhoff2021,
   author = {J Benkhoff and G Murakami and W Baumjohann and S Besse and E Bunce and M Casale and G Cremosese and K.-H. Glassmeier and H Hayakawa and D Heyner and H Hiesinger and J Huovelin and H Hussmann and V Iafolla and L Iess and Y Kasaba and M Kobayashi and A Milillo and I G Mitrofanov and E Montagnon and M Novara and S Orsini and E Quemerais and U Reininghaus and Y Saito and F Santoli and D Stramaccioni and O Sutherland and N Thomas and I Yoshikawa and J Zender},
   doi = {10.1007/s11214-021-00861-4},
   issn = {1572-9672},
   issue = {8},
   journal = {Space Science Reviews},
   pages = {90},
   title = {{BepiColombo - Mission Overview and Science Goals}},
   volume = {217},
   url = {https://doi.org/10.1007/s11214-021-00861-4},
   year = {2021}
}

@article{Bortis2008ActiveModulators,
    title = {{Active Gate Control for Current Balancing of Parallel-Connected IGBT Modules in Solid-State Modulators}},
    year = {2008},
    journal = {IEEE Transactions on Plasma Science},
    author = {Bortis, D and Biela, J and Kolar, J W},
    number = {5},
    pages = {2632--2637},
    volume = {36},
    doi = {10.1109/TPS.2008.2003971},
    issn = {1939-9375}
}

@inproceedings{Matallana2016AnalysisFundamentals,
    title = {{Analysis and modelling of IGBTs parallelization fundamentals}},
    year = {2016},
    booktitle = {IECON 2016 - 42nd Annual Conference of the IEEE Industrial Electronics Society},
    author = {Matallana, A and Andreu, J and Garate, J I and Aretxabaleta, I and Planas, E},
    number = {},
    pages = {3247--3252},
    volume = {},
    doi = {10.1109/IECON.2016.7793367}
}

@article{Zorngiebel2011ModularApplications,
    title = {{Modular 50-kV IGBT Switch for Pulsed-Power Applications}},
    year = {2011},
    journal = {IEEE Transactions on Plasma Science},
    author = {Zorngiebel, V and Hecquard, M and Spahn, E and Welleman, A and Scharnholz, S},
    number = {1},
    pages = {364--367},
    volume = {39},
    doi = {10.1109/TPS.2010.2068061},
    issn = {1939-9375}
}

@inproceedings{Anthon2014SwitchingPackage,
    title = {{Switching investigations on a SiC MOSFET in a TO-247 package}},
    year = {2014},
    booktitle = {IECON 2014 - 40th Annual Conference of the IEEE Industrial Electronics Society},
    author = {Anthon, A and Hernandez, J C and Zhang, Z and Andersen, M A E},
    number = {},
    pages = {1854--1860},
    volume = {},
    isbn = {1553-572X},
    doi = {10.1109/IECON.2014.7048754}
}

@article{Lev2019ThePropulsion,
    title = {{The technological and commercial expansion of electric propulsion}},
    year = {2019},
    journal = {Acta Astronautica},
    author = {Lev, Dan and Myers, Roger M and Lemmer, Kristina M and Kolbeck, Jonathan and Koizumi, Hiroyuki and Polzin, Kurt},
    pages = {213--227},
    volume = {159},
    url = {https://www.sciencedirect.com/science/article/pii/S0094576518319672},
    doi = {https://doi.org/10.1016/j.actaastro.2019.03.058},
    issn = {0094-5765}
}

@article{Chabert2012GlobalCoil,
    title = {{Global model of a gridded-ion thruster powered by a radiofrequency inductive coil}},
    year = {2012},
    journal = {Physics of Plasmas},
    author = {Chabert, P and Arancibia Monreal, J and Bredin, J and Popelier, L and Aanesland, A},
    number = {7},
    month = {7},
    pages = {073512},
    volume = {19},
    url = {https://doi.org/10.1063/1.4737114},
    doi = {10.1063/1.4737114},
    issn = {1070-664X}
}

@article{Taccogna2023PlasmaAlgorithms,
    title = {{Plasma propulsion modeling with particle-based algorithms}},
    year = {2023},
    journal = {Journal of Applied Physics},
    author = {Taccogna, F and Cichocki, F and Eremin, D and Fubiani, G and Garrigues, L},
    number = {15},
    month = {10},
    pages = {150901},
    volume = {134},
    url = {https://doi.org/10.1063/5.0153862},
    doi = {10.1063/5.0153862},
    issn = {0021-8979}
}

@article{Gallimore1993AnodeThrusters,
    title = {{Anode power deposition in magnetoplasmadynamic thrusters}},
    year = {1993},
    journal = {Journal of Propulsion and Power},
    author = {Gallimore, A D and Kelly, A J and Jahn, R G},
    number = {3},
    month = {5},
    pages = {361--368},
    volume = {9},
    publisher = {American Institute of Aeronautics and Astronautics},
    url = {https://doi.org/10.2514/3.23630},
    doi = {10.2514/3.23630},
    issn = {0748-4658}
}

@article{Zuin2004CriticalThruster,
    title = {{Critical regimes and magnetohydrodynamic instabilities in a magneto-plasma-dynamic thruster}},
    year = {2004},
    journal = {Physics of Plasmas},
    author = {Zuin, M and Cavazzana, R and Martines, E and Serianni, G and Antoni, V and Bagatin, M and Andrenucci, M and Paganucci, F and Rossetti, P},
    number = {10},
    month = {10},
    pages = {4761--4770},
    volume = {11},
    url = {https://doi.org/10.1063/1.1786593},
    doi = {10.1063/1.1786593},
    issn = {1070-664X}
}

@article{Choueiri1998ScalingThrusters,
    title = {{Scaling of Thrust in Self-Field Magnetoplasmadynamic Thrusters}},
    year = {1998},
    journal = {Journal of Propulsion and Power},
    author = {Choueiri, Edgar},
    number = {5},
    month = {9},
    pages = {744--753},
    volume = {14},
    publisher = {American Institute of Aeronautics and Astronautics},
    url = {https://doi.org/10.2514/2.5337},
    doi = {10.2514/2.5337},
    issn = {0748-4658}
}

@article{Krulle1998TechnologyPropulsion,
    title = {{Technology and Application Aspects of Applied Field Magnetoplasmadynamic Propulsion}},
    year = {1998},
    journal = {Journal of Propulsion and Power},
    author = {Krulle, Gerd and Auweter-Kurtz, Monika and Sasoh, Akihiro},
    number = {5},
    month = {9},
    pages = {754--763},
    volume = {14},
    publisher = {American Institute of Aeronautics and Astronautics},
    url = {https://doi.org/10.2514/2.5338},
    doi = {10.2514/2.5338},
    issn = {0748-4658}
}

@inproceedings{Woodley2026ConnectingThrusters,
  author    = {Woodley, Adrian and Heiser, Ryan and Underwood, Thomas C.},
  title     = {Connecting Ablative and Gas-Fed Propellant Utilization in Electromagnetic Thrusters},
  booktitle = {AIAA SCITECH 2026 Forum},
  series    = {AIAA SciTech Forum},
  publisher = {American Institute of Aeronautics and Astronautics},
  address   = {Reston, VA},
  month     = jan,
  year      = {2026},
  doi       = {10.2514/6.2026-0279},
}

@article{Subramaniam2018ComputationalAccelerators,
    title = {{Computational and experimental investigation of plasma deflagration jets and detonation shocks in coaxial plasma accelerators}},
    year = {2018},
    journal = {Plasma Sources Science Technology},
    author = {Subramaniam, Vivek and Underwood, Thomas C. and Raja, Laxminarayan L. and Cappelli, Mark A.},
    month = {2},
    volume = {27}
}

@article{Subramaniam2018AAccelerators,
    title = {{A plasma–vacuum interface tracking algorithm for magnetohydrodynamic simulations of coaxial plasma accelerators}},
    year = {2018},
    journal = {Journal of Computational Physics},
    author = {Subramaniam, Vivek and Raja, Laxminarayan L},
    pages = {207--225},
    volume = {366},
    url = {https://www.sciencedirect.com/science/article/pii/S0021999118302079},
    doi = {https://doi.org/10.1016/j.jcp.2018.03.041},
    issn = {0021-9991}
}

@article{Ziemer2001ScalingThrusters,
    title = {{Scaling laws for electromagnetic pulsed plasma thrusters}},
    year = {2001},
    journal = {Plasma Sources Science and Technology},
    author = {Ziemer, J K and Choueiri, E Y},
    number = {3},
    pages = {395--405},
    volume = {10},
    publisher = {},
    url = {https://doi.org/10.1088/0963-0252/10/3/302},
    doi = {10.1088/0963-0252/10/3/302},
    issn = {0963-0252}
}

@article{Cohen1950TheGas,
    title = {{The Electrical Conductivity of an Ionized Gas}},
    year = {1950},
    journal = {Physical Review},
    author = {Cohen, Robert S and Spitzer, Lyman and Routly, Paul McR.},
    number = {2},
    month = {10},
    pages = {230--238},
    volume = {80},
    publisher = {American Physical Society},
    url = {https://link.aps.org/doi/10.1103/PhysRev.80.230},
    doi = {10.1103/PhysRev.80.230}
}

@article{Baghirzade2026MHDVLEO,
    title = {{MHD modeling of magneto-deflagration and magneto-detonation modes of air plasma jets in coaxial plasma accelerators in VLEO}},
    year = {2026},
    journal = {Journal of Applied Physics},
    author = {Baghirzade, Mammadbaghir and Underwood, Thomas C and Raja, Laxminarayan L},
    number = {4},
    month = {7},
    pages = {043302},
    volume = {140},
    url = {https://doi.org/10.1063/5.0325672},
    doi = {10.1063/5.0325672},
    issn = {0021-8979}
}

@article{Liu2024EffectsMode,
    title = {{Effects of discharge parameters on plasma acceleration and transmission characteristics of a coaxial gun operated in gas-prefilled mode}},
    year = {2024},
    journal = {Journal of Applied Physics},
    author = {Liu, Shuai and Qi, Liangwen and Zhang, Guipeng and Xiao, Dingbang and Yu, Siqi},
    number = {16},
    month = {10},
    pages = {163301},
    volume = {136},
    url = {https://doi.org/10.1063/5.0229983},
    doi = {10.1063/5.0229983},
    issn = {0021-8979}
}

@article{Crandall2022Air-breathingAnalysis,
    title = {{Air-breathing electric propulsion: mission characterization and design analysis}},
    year = {2022},
    journal = {Journal of Electric Propulsion},
    author = {Crandall, Patrick and Wirz, Richard E},
    number = {1},
    pages = {12},
    volume = {1},
    url = {https://doi.org/10.1007/s44205-022-00009-8},
    doi = {10.1007/s44205-022-00009-8},
    issn = {2731-4596}
}

\end{document}